\documentclass[aps,twocolumn,pra,groupedaddress,superscriptaddress,amsmath,amssymb,amsthm]{revtex4}
\usepackage{subfigure,hyperref,bbm,times}
\usepackage[T1]{fontenc}
\usepackage{braket}
\usepackage{amsthm}
\usepackage{epsfig}
\usepackage{color}
\usepackage{graphicx}
\usepackage{dcolumn}
\usepackage{bm}

\providecommand{\openone}{\leavevmode\hbox{\small1\kern-3.8pt\normalsize1}}

\usepackage{soul}

\begin{document}

\title{Activating remote entanglement in a quantum network by local counting of identical particles}

\author{Alessia Castellini}
\affiliation{Dipartimento di Fisica e Chimica - Emilio Segr\`e, Universit\`a di Palermo, via Archirafi 36, 90123 Palermo, Italy}

\author{Bruno Bellomo}
\affiliation{Institut UTINAM - UMR 6213, CNRS, Universit\'{e} Bourgogne Franche-Comt\'{e}, Observatoire des Sciences de l'Univers THETA, 41 bis avenue de l'Observatoire, F-25010 Besan\c{c}on, France}

\author{Giuseppe Compagno}
\affiliation{Dipartimento di Fisica e Chimica - Emilio Segr\`e, Universit\`a di Palermo, via Archirafi 36, 90123 Palermo, Italy}

\author{Rosario Lo Franco}
\email{rosario.lofranco@unipa.it}
\affiliation{Dipartimento di Fisica e Chimica - Emilio Segr\`e, Universit\`a di Palermo, via Archirafi 36, 90123 Palermo, Italy}
\affiliation{Dipartimento di Ingegneria, Universit\`{a} di Palermo, Viale delle Scienze, Edificio 6, 90128 Palermo, Italy}

\date{\today }

\begin{abstract}
Quantum information and communication processing within quantum networks usually employs identical particles. Despite this, the physical role of quantum statistical nature of particles in large-scale networks remains elusive. Here, we show that just the indistinguishability of fermions makes it possible a new mechanism of entanglement transfer in many-node quantum networks. This process activates remote entanglement among distant sites, which do not share a common past, by only locally counting identical particles and classical communication. These results constitute the key achievement of the present technique and open the way to a more stable multistage transfer of nonlocal quantum correlations based on fermions.

\end{abstract}



\maketitle
\section{Introduction}
New avenues have recently been opened in quantum information and communication by the transfer of quantum state among different separated sites \cite{quantumRepeaters}. In fact, it allows distributed quantum computing, a quantum internet and tests of quantum mechanics foundations within composite quantum networks \cite{castelvecchiQuantInternet,pirandolaQuantInternet,humphreys2018deterministic}. Usual state transfer procedures employ identical particles (that are elementary subsystems such as atoms, photons, electrons, qubits), where their entanglement plays an essential role and no effect associated to the statistical nature of the particles typically shows up. This is due to the fact that, in these processes, spatial overlap of the wave functions does not occur in the relevant places so that the identical particles are distinguishable and behave like non-identical ones. One may then inquire whether employing identical particles may lead to new features in the context of quantum communication by exploiting  indistinguishability as a direct resource. For this to happen, it is required to investigate those situations where particles spatially overlap, so that particle identity implies their indistinguishability.
Identity of particles has been shown to be useful for some quantum information protocols \citep{omarIJQI,PhysRevA.65.062305,PhysRevLett.88.187903,PhysRevA.68.052309,bose2002indisting,LFCSciRep,sciaraSchmidt,bellomo2017,compagno2018dealing,
LoFrancoPRL} and for quantum metrology \cite{benatti2014NJP, QEMreview}. In this context, one of the problems which remains to investigate is the role played by the quantum statistical nature of identical particles into the mechanisms of entanglement transfer within large-scale networks.

Among the various mechanisms of many-node state transfer, we focus on the entanglement swapping (ES), which is a must for large-scale distribution of quantum information \citep{quantcomm,quantcomm2,quantcomm3,sun2017entanglement} and is subject of intense experimental interest \citep{pan1998experimental,experiment,yang2006experimental,exp3,megidish2013,2017timebin}.
ES is an intrinsically quantum phenomenon which permits to entangle two particles not sharing a common past, each particle being outside the light cone of the other. ES constitutes a key process to implement quantum communication \citep{quantcomm,quantcomm2,quantcomm3,sun2017entanglement} and is crucial to build quantum relays and quantum repeaters \citep{quantumRelay1,quantumRepeaters}. So far, in all the implementations, essential ingredients are the initial creation of entangled pairs and Bell measurements \citep{BSA}. In the standard ES process \citep{1zuk1993}, two entangled particle pairs are initially prepared and a Bell measurement is successively performed on two particles of the different pairs. As a result, the other two particles become entangled even if they never interacted \citep{nonloc,nonlocality}. ES has been experimentally realized using identical but distinguishable particles (photons) by applying the usual operational framework for non-identical particles, based on particle addressability (local operations and classical communication (LOCC)) \cite{horodecki2009quantum}. The initial entangled pairs of photons are typically created by spontaneous parametric down conversion (SPDC) \citep{pan1998experimental,experiment,megidish2013,yang2006experimental,exp3}.
Recently, ES has been successfully achieved in a quantum network, entangling two photons over a distance of 100 km \citep{2017timebin}. Multiple ES has been theoretically proposed and experimentally realized by extension of the standard protocol \citep{bose1998multiparticle,2008multistage,2013multistage,MultipartEnt2009}.
The overall success of the process is influenced by the low creation rate of photon pairs in SPDC \citep{SPDC,SPDC2,problemsSPDC} and by the inefficiency in the realization of Bell measurements \citep{BSA,yang2006experimental,lee2013bell,cQED,cQED3,PhysRevA.84.042331,PhysRevA.92.042314,PhysRevA.82.032318}.

In this work we present a new process of entanglement transfer in a many-node quantum network, where neither initial entangled particle pairs nor Bell measurements are needed, exploiting indistinguishability of fermions. Although the idea of using indistinguishability of identical particles to generate entanglement is not a new one, the present process shows that remote entanglement among distant sites can be generated by only locally counting particles. These characteristics constitute the key achievement of the present technique.
In the presence of spatial overlap, the identity of particles makes them individually unaddressable. Therefore, we employ an approach based on spatially localized operations and classical communication (sLOCC), where single-particle local measurements are made onto assigned spatial regions \citep{LoFrancoPRL,compagno2018dealing} (closer to the spirit of quantum field theory). We finally compare this process to that with bosons.

\begin{figure}[!t]
\centering
\includegraphics[width=0.44 \textwidth]{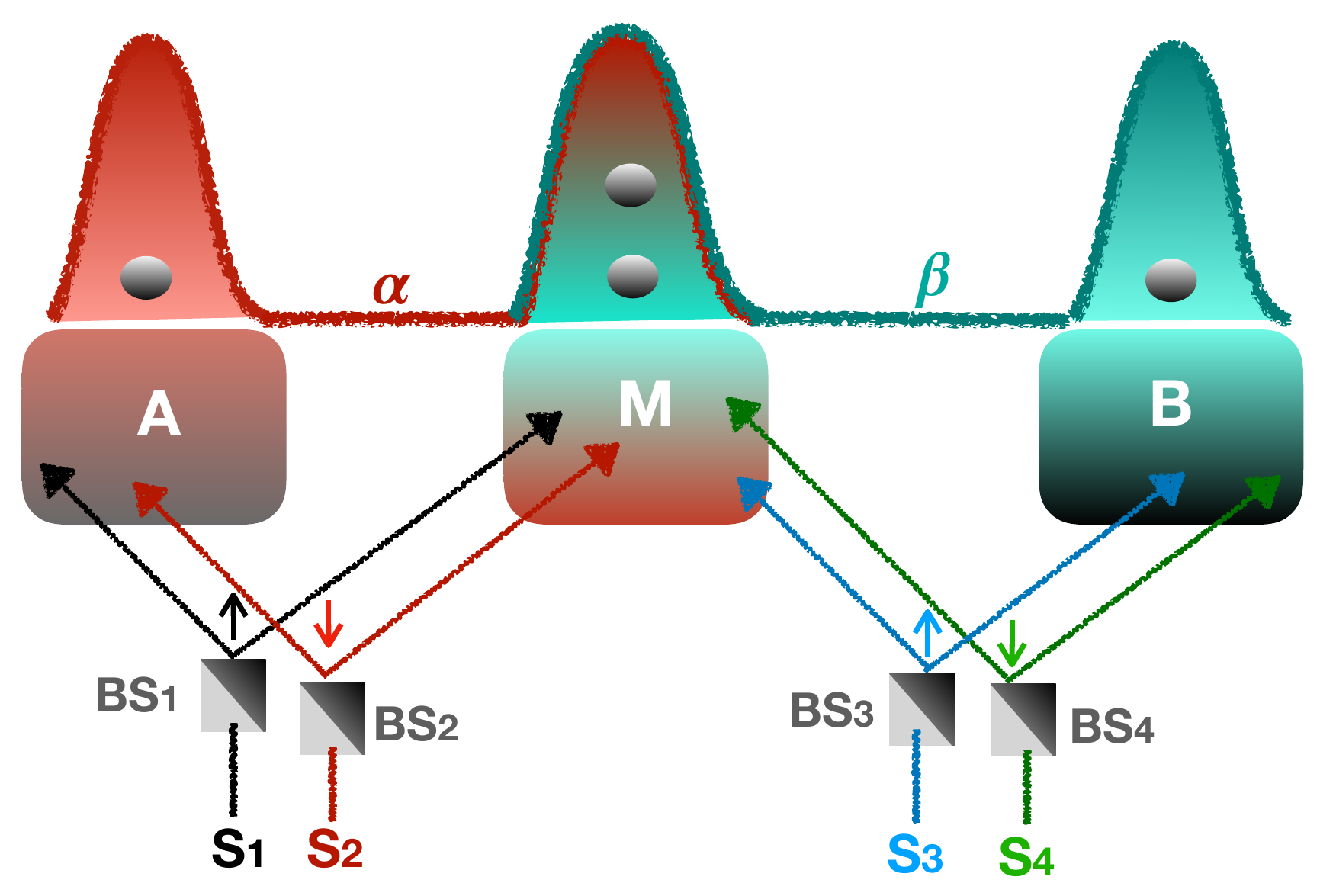}
\caption{\textbf{Basic process.} Scheme for the entanglement transfer by four independently-prepared indistinguishable fermions, with a shared intermediate node $\mathrm{M}$. The delocalized spatial modes $\alpha$ and $\beta$ partially overlap in correspondence of the intermediate node $\mathrm{M}$. Post-selection by sLOCC leaves two fermions with opposite pseudospins in the central node M and two entangled fermions in the extreme nodes, A and B.}
\label{Fig1}
\end{figure}

\section{Basic process with fermions}
Experimental techniques to control fermions in quantum networks have been recently developed \citep{macroscES2006,Yamamoto2008,feve2007demand,bocquillon2013coherence,PhysRevLett.121.166801,FermionicStatistics,sansoni2012two}. In the following we focus on unveiling remarkable aspects introduced by both fermionic statistics and spatial overlap in the process of remote entanglement distribution.

To this aim, we first describe the basic process which serves as the elementary step for the extension to a many-node quantum network.
As displayed in Fig.~\ref{Fig1}, we take four identical fermions (two pairs), prepared by four independent (space-like separated) sources $\{\mathrm{S}_i, \ i=1,...,4\}$. Each particle is sent to the corresponding beam splitter $\mathrm{BS}_i$. The two sources $\mathrm{S}_1$ and $\mathrm{S}_2$ independently prepare two fermions with opposite pseudospin. Each beam splitter sends the particle with the same amplitude into two separated sites A and M, so that each particle is in the same delocalized spatial mode $\ket{\alpha}=(\ket{\mathrm{A}}+\ket{\mathrm{M}})/\sqrt{2}$.
Similarly, sources $\mathrm{S}_3$ and $\mathrm{S}_4$ generate the fermions of the second pair with opposite pseudospin in the delocalized spatial mode
$\ket{\beta}=(\ket{\mathrm{M}}+\ket{\mathrm{B}})/\sqrt{2}$ (right side of Fig.~\ref{Fig1}). The modes $\ket{\alpha}$ and $\ket{\beta}$ partially overlap in the shared intermediate node M and the I-th node (I $=$ A, M, B) is chosen such that only the localized bound state $\ket{\mathrm{I}}$ is present. The condition that particles are in the chosen localized bound states can be assured by keeping only the cases when local detectors do not measure particles elsewhere and by classically communicating the results. The initially prepared four-fermion state $|\Psi^{(4)}_{\mathrm{f}}\rangle$ can be then formally obtained from $\ket{\alpha \downarrow,\alpha \uparrow,\beta \downarrow,\beta \uparrow}$ by dropping, because of the Pauli exclusion principle, the terms with the same pseudospins in the central node (same spatial state $|\mathrm{M}\rangle$) \cite{orzelPRA}. The ultimate scope is to generate entanglement between particles in the far nodes $\mathrm{A}$ and $\mathrm{B}$. This can be achieved by using sLOCC \citep{LoFrancoPRL}, which here consist in a post-selection locally counting only one particle of the first pair in A and one particle of the second pair in B and using classical communication among these sites (this counting implies that in the central node M there are two particles). The classical communication allows sites A and B to know when an entangled pair is obtained. Notice that the local counting operation is a free operation with respect to entanglement \cite{plenio2007mb}. Such a post-selection can be implemented utilizing, for instance, absorbtionless particle-counting detectors in A and B, which do not disturb the pseudospin state \citep{detectorselectrons,detectors,detectors2,bose2002indisting}.
Similar non-demolition measurements are applied in the other post-selections by sLOCC used along the paper.

One then gets the post-selected global state (see Appendix \ref{AppF})
\begin{equation}\label{Psifent2}
|\Psi^{(4)}_{\mathrm{f,PS}}\rangle=|\Psi_\mathrm{M}, \Psi_{\mathrm{AB}}^{-}\rangle,
\end{equation}
where
\begin{equation}\label{Psi2AB}
\ket{\Psi_\mathrm{M}}=\ket{\mathrm{M}\uparrow,\mathrm{M}\downarrow},\ \ket{\Psi_{\mathrm{AB}}^{-}} = \dfrac{|\mathrm{A}\downarrow,\mathrm{B}\uparrow\rangle-|\mathrm{A}\uparrow,\mathrm{B}\downarrow\rangle}{\sqrt{2}}.
\end{equation}
We have thus obtained a maximally entangled state of two-particle pseudospins over the distant nodes $\mathrm{A}$ and $\mathrm{B}$, despite the latter are always independent and the particles do not share any common past. We stress that if the four particles are not identical, the same post-selection procedure does not give rise to an entangled state.
The state of Eq.~\eqref{Psifent2} is obtained with probability (see Appendix \ref{AppA}) $P_{\mathrm{f}}(4)=|\langle \Psi_{\mathrm{f,PS}}^{(4)}|\Psi_{\mathrm{f}}^{(4)}\rangle|^2=\langle \Psi_{\mathrm{f}}^{(4)}|\hat{\Pi}_{\mathrm{f}}|\Psi_{\mathrm{f}}^{(4)}\rangle=2/9$, where  $\hat{\Pi}_\mathrm{f}=\sum_{\sigma,\tau=\uparrow,\downarrow}|\mathrm{A} \ \sigma,\mathrm{M}\uparrow,$ $\mathrm{M}\downarrow,  \mathrm{B}\ \tau \rangle\langle \mathrm{A}\ \sigma,\mathrm{M}\uparrow,\mathrm{M}\downarrow, \mathrm{B}\ \tau|$ is the projector onto the (AMB)-operational subspace. We remark that this entanglement distribution is reached without entanglement-inducing Bell measurements on the central particles, but just exploiting the indistinguishability of non-interacting fermions in M. In fact, the spatial overlap of fermions in the shared intermediate site $\mathrm{M}$ plays the key role of an entanglement-transfer gate. Schemes for entanglement swapping without Bell measurements have been proposed in contexts where interaction is essential (e.g., cavity QED) \citep{cQED,cQED2,cQED3,CQED2017}.

\begin{figure*}[t!]
\begin{center}
\includegraphics[scale=0.44]{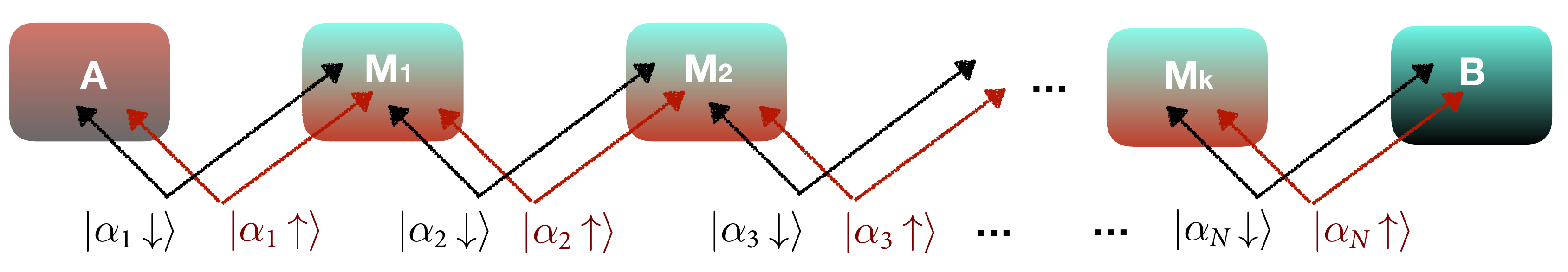}
\caption{\textbf{Many-node quantum network.} Scheme for remote entanglement distribution between distant nodes A and B by $n$ independently-prepared indistinguishable particles with $k=N-1$ shared intermediate nodes, being $N=n/2$ the number of involved particle pairs. This process is the straightforward generalization of the scheme of Fig.~\ref{Fig1}.}
\label{Fig2}
\end{center}
\end{figure*}

\section{Large-scale process with fermions}
The basic scheme of Fig.~\ref{Fig1} can be straightforwardly iterated to create a remote entanglement transfer in a many-node quantum network. This is achieved by means of $n$ identical fermions and $k=N-1$ shared intermediate nodes $\mathrm{M}_i$ ($i=1,\ldots,k$), where $N=n/2$ is the number of particle pairs. As displayed in Fig.~\ref{Fig2}, each $j$-th pair ($j=1,\ldots,N$) has opposite pseudospins and spatial mode $\ket{\alpha_j}$, with $\ket{\alpha_1} = (\ket{\mathrm{A}}+\ket{\mathrm{M}_1})/\sqrt{2}$, $\ket{\alpha_N} = (\ket{\mathrm{M}_k}+\ket{\mathrm{B}})/\sqrt{2}$ and $\ket{\alpha_j} = (\ket{\mathrm{M}_{j-1}}+\ket{\mathrm{M}_{j}})/\sqrt{2}$ for $j=2,\ldots,k$. The aim is to activate entanglement of particles in the remote far nodes A and B of the network.
The modes $\ket{\alpha_i}$ and $\ket{\alpha_{i+1}}$ partially overlap in the shared intermediate node $\mathrm{M}_i$ and the I-th node (I $=$ A, M$_i$, B) is taken, as already mentioned above, such that only the localized bound state $\ket{\mathrm{I}}$ is present.
Once again the initial $n$-fermion state $|\Psi^{(n)}_{\mathrm{f}}\rangle$ can be formally determined starting from the state $\ket{\alpha_1\downarrow, \alpha_1\uparrow,\ldots,\alpha_N\downarrow,\alpha_N\uparrow}$ simply by dropping, because of the Pauli exclusion principle, the terms having the same pseudospins in each intermediate node. After that, by counting one particle in A and one in B (this entails that each node $\mathrm{M_i}$ contains two particles) and allowing for classical communication as before, the post-selected global state is
\begin{equation}
|\Psi^{(n)}_{\mathrm{f,PS}}\rangle=
|\Psi_{\mathrm{M}_1},\Psi_{\mathrm{M}_2},\ldots,\Psi_{\mathrm{M}_k}, \Psi_{\mathrm{AB}}^{-}\rangle,
\end{equation}
where $\ket{\Psi_{\mathrm{M}_i}}=\ket{\mathrm{M}_i\uparrow,\mathrm{M}_i\downarrow}$ and $\ket{\Psi_{\mathrm{AB}}^{-}}$ is the maximally entangled (Bell) state of Eq.~\eqref{Psi2AB}. The probability to obtain the state above is $P_{\mathrm{f}}(n)=|\langle \Psi_{\mathrm{f,PS}}^{(n)}|\Psi_{\mathrm{f}}^{(n)}\rangle|^2$ (see Appendix \ref{AppF} for its explicit expression).
Thus, we have generated entanglement of particle pseudospins between the independent distant locations A and B of the many-node network, starting with independently-prepared identical fermions, with no Bell measurements and only using local counting operations.
We remark that all these features make the remote entanglement activation based on identical fermions deeply different from the standard processes of entanglement transfer such as ES.

\section{Process with bosons} 
The basic setup of Fig.~\ref{Fig1} can be also thought to be run by identical bosons. We shall show that, in this case, a Bell measurement onto the intermediate site M is eventually required for achieving the desired entanglement transfer, similarly to a standard protocol of ES.
The sLOCC framework is now realized by locally counting two particles in the intermediate node M and only one particle in each of the far nodes A and B, also allowing for classical communication among the different sites. From the initial (unnormalized) state $\ket{\alpha \downarrow,\alpha \uparrow,\beta \downarrow,\beta \uparrow}$, one gets the four-boson post-selected state (see Appendix \ref{AppB})
\begin{equation}\label{finalbos}
|\Psi_{\mathrm{b,PS}}^{(4)}\rangle=
\frac{\ket{\Psi_{\mathrm{M}},\Psi^{+}_{\mathrm{AB}}}
+\ket{\Phi^{+}_{\mathrm{M}},\Phi^{+}_{\mathrm{AB}}}
- \ket{\Phi^{-}_{\mathrm{M}},\Phi^{-}_{\mathrm{AB}}}}{\sqrt{3}},
\end{equation}
where $|\Phi^{\pm}_{\mathrm{M}}\rangle=(|\mathrm{M}\downarrow,\mathrm{M}\downarrow\rangle\pm|\mathrm{M}\uparrow,\mathrm{M}\uparrow\rangle)/2$, $\ket{\Psi_{\mathrm{M}}}$ is given in Eq.~\eqref{Psi2AB}, while the distant sites A and B share the Bell states
\begin{align}
\begin{split}
&|\Psi^+_{\mathrm{AB}}\rangle=\dfrac{1}{\sqrt{2}}(|\mathrm{A\downarrow,B\uparrow}\rangle+|\mathrm{A\uparrow,B\downarrow}\rangle),\\
&|\Phi^\pm_{\mathrm{AB}}\rangle=\dfrac{1}{\sqrt{2}}(|\mathrm{A\downarrow,B\downarrow}\rangle\pm |\mathrm{A\uparrow,B\uparrow}\rangle).
\end{split}
\end{align}
The presence of these three Bell states is a consequence of the fact that bosonic systems admit two-particle states with the same pseudospins in M.
As in the standard ES procedure, a joint (Bell) measurement in the shared intermediate node $\mathrm{M}$ determines the entangled state in which the first and the last boson of the network collapse, each outcome occurring with probability $p=1/3$, as seen from Eq. \eqref{finalbos}. Since each of the three Bell-state outcomes from the joint measurement in M realizes the desired entanglement transfer over A and B, the success probability of the process coincides with the probability of obtaining the post-selected state of Eq.~\eqref{finalbos}, which is $P_{\mathrm{b}}(4)=|\langle \Psi_{\mathrm{b,PS}}^{(4)}|\Psi_{\mathrm{b}}^{(4)}\rangle|^2=6/25$.
This bosonic protocol can be then extended, analogously to the standard multiple ES, by a cascaded procedure \cite{2008multistage}. The scheme remains that of Fig.~\ref{Fig2} with $n$ independently-prepared identical bosons and $k=N-1$ intermediate nodes ($N=n/2$). The sLOCC framework again consists in counting two particles in each intermediate node and one in the distant nodes A and B, also allowing for classical communication of the counting outcomes. One gets the post-selected state
$\ket{\Psi_{\mathrm{b,PS}}^{(n)}}$ and performs Bell measurements step by step on each intermediate node $\mathrm{M}_i$ ($i=1,...,k$) to transfer entanglement over A and B. The success probability is $P_{\mathrm{b}}(n)=|\langle \Psi_{\mathrm{b,PS}}^{(n)}|\Psi_{\mathrm{b}}^{(n)}\rangle|^2$ (see Appendix \ref{AppB}).

\section{Process with separated intermediate sites}
\begin{figure}[!t]
\centering
\includegraphics[width=0.46 \textwidth]{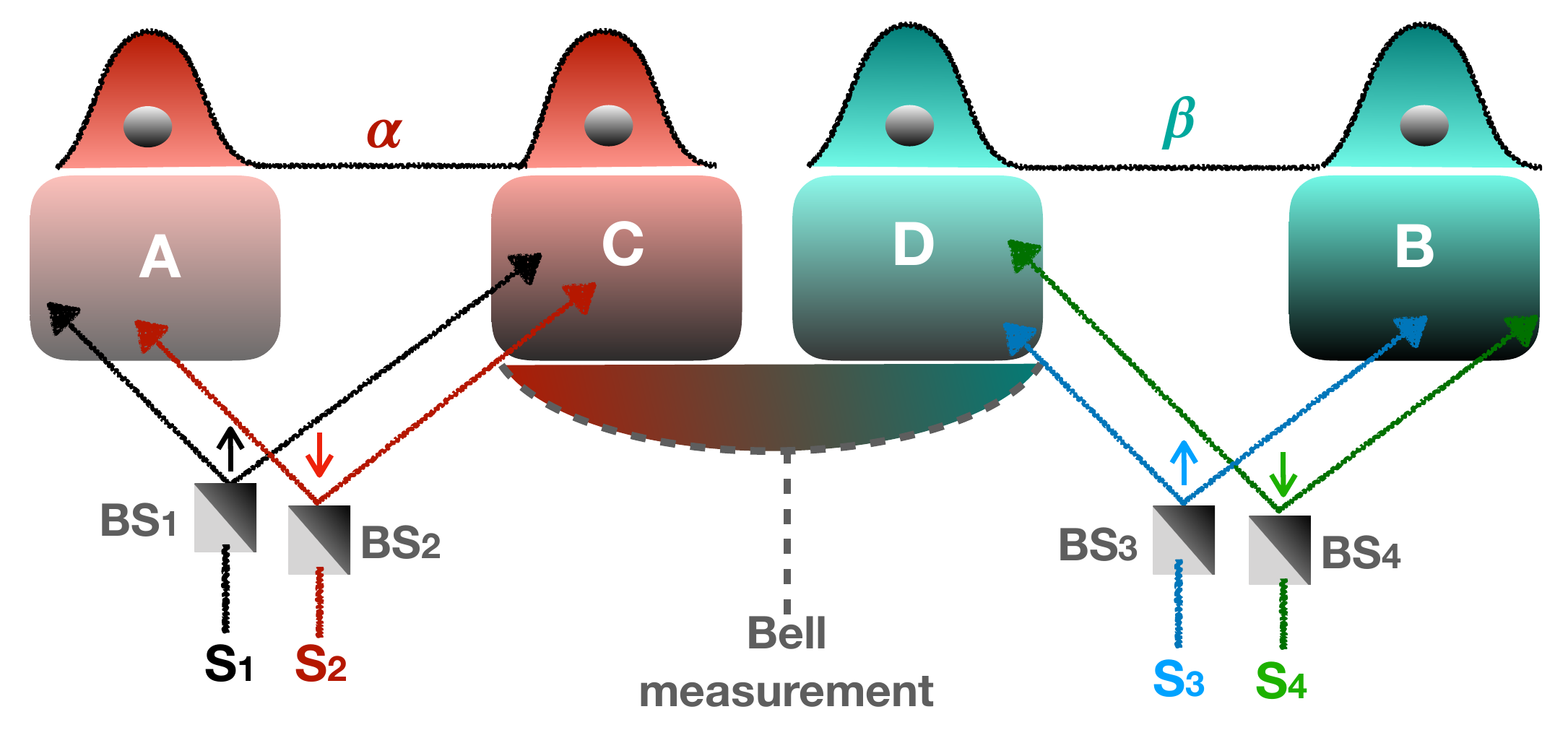}
\caption{Four-node scheme for the entanglement swapping by indistinguishable particles (bosons or fermions). Four independent single-particle sources $\mathrm{S}_i$ ($i=1,...,4$) send each particle to the corresponding beam splitter ($\mathrm{BS}_i$). $\alpha$ and $\beta$ are the (delocalized) spatial modes peaked in correspondence of the separated spatial nodes A-C and D-B, respectively. Post-selection by sLOCC leaves only one particle in each node and Bell measurements are finally performed.}
\label{Fig3}
\end{figure}

Notice that a procedure much closer to the standard ES can be moreover obtained with indistinguishable particles (bosons or fermions) by employing intermediate separated nodes instead of common intermediate ones. 

We take a system made of four identical particles (either bosons or fermions), prepared by four independent (space-like separated) sources $\{\mathrm{S}_i, \ i=1,...,4\}$. Each particle is sent to the corresponding beam splitter $\mathrm{BS}_i$, as depicted in Fig.~\ref{Fig3}. The two sources $\mathrm{S}_1$ and $\mathrm{S}_2$ independently prepare two particles with opposite pseudospin. Each beam splitter sends the particle with the same amplitude into two separated sites A and C, so that each particle is in the same delocalized spatial mode $|\alpha\rangle=(|\mathrm{A}\rangle+|\mathrm{C}\rangle)/\sqrt{2}$.
Similarly, sources $\mathrm{S}_3$ and $\mathrm{S}_4$ generate the particles of the second pair with opposite pseudospin in the delocalized spatial mode
$|\beta\rangle=(|\mathrm{D}\rangle+|\mathrm{B} \rangle)/\sqrt{2}$ (right side of Fig.~\ref{Fig3}). All the nodes are spatially separated (the modes $\ket{\alpha}$ and $\ket{\beta}$ are orthogonal) and the I-th node (I $=$ A, B, C, D) is chosen such that only the localized bound state $\ket{\mathrm{I}}$ is present. The condition that particles are in the chosen localized bound states can be assured by keeping only the cases when local detectors do not measure particles elsewhere and by classically communicating the results.
The global four-particle quantum state \citep{LFCSciRep} is therefore $|\Psi^{(4)}\rangle=|\alpha \downarrow,\alpha \uparrow,\beta \downarrow, \beta \uparrow\rangle$.
From this state it is possible to obtain entanglement in the pseudospin degrees of freedom linked to the spatial overlap of particles in each pair. This is achieved by sLOCC \citep{LoFrancoPRL}, which here consist in a post-selection counting only one particle of the first pair in A and one particle of the second pair in B and classically communicating this outcome to each other. This post-selection can be implemented using, for instance, one absorbtionless particle-counting detector in A and one in B, which do not disturb the pseudospin state \citep{detectorselectrons,detectors,detectors2,bose2002indisting}.
Similar non-demolition measurements are applied in the other post-selections used along the paper. As a result, each node contains one particle and we obtain the state (see Appendix \ref{AppSeparated})
\begin{equation}\label{PsiPStensor}
\ket{\Psi_{\mathrm{PS}}^{(4)}} = \ket{\Psi_{\mathrm{AC}},\Psi_{\mathrm{DB}}},
\end{equation}
where $\ket{\Psi_{\mathrm{AC}}}$ and $\ket{\Psi_{\mathrm{DB}}}$ are the two-particle Bell states
\begin{eqnarray}\label{BellStates}
\ket{\Psi_{\mathrm{AC}}} = \dfrac{1}{\sqrt{2}}(|\mathrm{A}\downarrow,\mathrm{C} \uparrow\rangle+\eta|\mathrm{A}\uparrow,\mathrm{C} \downarrow\rangle), \nonumber\\
\ket{\Psi_{\mathrm{DB}}} = \dfrac{1}{\sqrt{2}}(|\mathrm{D}\downarrow,\mathrm{B}\uparrow\rangle+\eta|\mathrm{D}\uparrow,\mathrm{B} \downarrow\rangle).
\end{eqnarray}
Even if the particles have been independently prepared, as a consequence of sLOCC, the state $\ket{\Psi_{\mathrm{PS}}^{(4)}}$ shows that the pair of particles in $\mathrm{A}$ and $\mathrm{C}$ is maximally entangled in the pseudospin degrees of freedom, as the $\mathrm{DB}$-pair.
This state is obtained with probability $P(4)=|\langle \Psi_{\mathrm{PS}}^{(4)}|\Psi^{(4)}\rangle|^2=1/4$. At this stage the particles can be distinguished, since they are in spatially separated sites. We stress that for each pair, if the particles are not identical, the same post-selection procedure  does not give rise to an entangled state. The structure of the state of Eq.~\eqref{PsiPStensor} allows to implement the standard protocol of entanglement swapping (ES) \citep{1zuk1993}: performing a Bell measurement on near central nodes C and D transfers entanglement to the particles in the far nodes A and B. Notice that this procedure does not require, at the preparation stage, two entangled pairs. The present scheme works for both bosons and fermions, also when particles of different pairs are not identical.
Moreover, in analogy with the standard ES, it can be naturally iterated by a cascaded procedure \cite{2008multistage} to realize multistage entanglement swapping with $n=2N$ independently-prepared particles, being $N$ the number of involved particle pairs. This is achieved by using a network with $n-2$ separated central nodes, where each pair of identical particles (either bosons or fermions) is prepared with opposite pseudospins in an equal delocalized spatial mode peaked in correspondence of two separated nodes (as shown for the two pairs in Fig.~\ref{Fig3}). After obtaining a single particle in each central node and performing Bell measurements step by step onto two central nodes \cite{2008multistage}, one eventually entangles the particles in the extreme nodes of the network with probability $P(n)=1/2^{N}$.

\section{Conclusions}
In this work we have presented a new conceptual process of entanglement distribution in a large-scale quantum network which is fundamentally activated by indistinguishability of particles.
The standard entanglement swapping, that is the renowned process for entanglement transfer with distinguishable particles, necessitates to start from entangled particle pairs and requires final Bell measurements \citep{1zuk1993}. Compared to this one, the present process, run by identical fermions, enables remote entanglement among distant nodes through the following different aspects: (i) with no distribution of initial entangled pairs and (ii) without performing Bell measurements, because of the natural entanglement due to spatially overlapping identical particles. Therefore, the process only requires local counting of independently-prepared identical particles.
The measurement process, when described on the level of particles, looks different for indistinguishable and distinguishable particles.

Besides its conceptual novelty, the key advantage of this process is that it simplifies the task of distributing entanglement, overcoming the drawbacks encountered in the usual entanglement transfer procedures during the initial preparation stage and the final measurement phase. In fact, it skips the use of sources of entangled particle pairs, which are for instance generated by SPDC at the very low rate of about $10^{-2}$ for single laser pulse \citep{SPDC2}, and also avoids the experimental inefficiency associated to performing Bell measurements \citep{BSA,yang2006experimental,lee2013bell,cQED,cQED3,PhysRevA.84.042331,PhysRevA.92.042314,PhysRevA.82.032318}.

The proposed fermionic process could be, for instance, realized by using quantum dots as sources of single electrons that can be initialized in particular spin states \cite{Yamamoto2008}, emitted on demand \cite{feve2007demand} and directed to quantum point contacts acting as electronic beam splitters \cite{bocquillon2013coherence,electronsReview}. Single electrons have been also recently shown to be controlled within atomic circuits \cite{PhysRevLett.121.166801}. Moreover, further setups in quantum optics, simulating fermionic statistics using photons and integrated photonics \cite{FermionicStatistics,sansoni2012two} could, in principle, represent convenient platforms.

Our results make it emerge once more \citep{LoFrancoPRL} that spatial overlap of identical particles constitutes an operational resource.
In addition, they pave the way to a more stable multistage remote entanglement transfer based on fermions, evidencing the effect of quantum statistical nature of particles on quantum information processing.

\textbf{Acknowledgments.} A.C. acknowledges for useful discussions Dario Cilluffo, Mauro Valeri, Andrea Geraldi and Emanuele Polino.

\appendix

\section{Explicit calculations for shared intermediate sites with fermions}\label{AppF}
The four-fermion global state $|\Psi^{(4)}_{\mathrm{f}}\rangle$ is
\begin{align}\label{ferm}
\begin{split}
|\Psi^{(4)}_{\mathrm{f}}\rangle=&\dfrac{1}{3}(|\mathrm{A\downarrow,A\uparrow,M\downarrow,M\uparrow\rangle+|A\downarrow,A\uparrow,M\downarrow,B\uparrow\rangle}\\
&+|\mathrm{A\downarrow,A\uparrow,B\downarrow,M\uparrow\rangle+|A\downarrow,A\uparrow,B\downarrow,B\uparrow\rangle}\\
&+|\mathrm{A\downarrow,M\uparrow,M\downarrow,B\uparrow\rangle}+|\mathrm{A\downarrow,M\uparrow,B\downarrow,B\uparrow\rangle}\\
&+|\mathrm{M\downarrow,A\uparrow,B\downarrow,M\uparrow\rangle+|M\downarrow,A\uparrow,B\downarrow,B\uparrow\rangle}\\
&+|\mathrm{M\downarrow,M\uparrow,B\downarrow,B\uparrow\rangle}).
\end{split}
\end{align}
The sLOCC here consists in counting one particle in A and one in B (this entails having two particles in M) and allowing for classical communication of the outcomes. Projecting therefore the above prepared state onto the subspace spanned by the basis $\mathcal{B}_{\mathrm{f}}=\{|\mathrm{A \ \sigma, M \uparrow, M \downarrow, B \tau \rangle}\}$ $(\sigma,\tau= \ \downarrow,\uparrow)$, we find the post-selected state
\begin{equation}
|\Psi^{(4)}_{\mathrm{f,PS}}\rangle=|\mathrm{M\uparrow,M\downarrow}\rangle \wedge \ket{\Psi_{\mathrm{AB}}^{-}},
\end{equation}
which is obtained with probability $P_{\mathrm{f}}(4)=|\langle \Psi^{(4)}_{\mathrm{f,PS}}|\Psi^{(4)}_{\mathrm{f}}\rangle|^2=2/9$. The state $\ket{\Psi_{\mathrm{AB}}^{-}}$ is the maximally Bell state of Eq.~\eqref{Bell}.

The scheme presented for the minimum core with four particles can be extended to the case of $n=2N$ particles, where $N$ is the number of involved particle pairs. The generalized scheme with $k=N-1$ shared intermediate nodes $\mathrm{M}_i$ ($i=1,\ldots,k$) is displayed in Fig. \ref{Fig1} of the main text.

Each $j$-th pair ($j=1,\ldots,N$) has opposite pseudospins and (delocalized) spatial mode $\ket{\alpha_j}$, with $\ket{\alpha_1} = (\ket{\mathrm{A}}+\ket{\mathrm{M}_1})/\sqrt{2}$, $\ket{\alpha_N} = (\ket{\mathrm{M}_k}+\ket{\mathrm{B}})/\sqrt{2}$ and $\ket{\alpha_j} = (\ket{\mathrm{M}_{j-1}}+\ket{\mathrm{M}_{j}})/\sqrt{2}$ for $j=2,\ldots,k$. We take as the initially prepared $n$-fermion state $|\Psi^{(n)}_{\mathrm{f}}\rangle$ the one obtained from $\ket{\alpha_1\downarrow, \alpha_1\uparrow,\ldots,\alpha_N\downarrow,\alpha_N\uparrow}$ by eliminating, because of the Pauli exclusion principle, the terms with two particles in the same node with the same pseudospin. The normalization constant $\mathcal{N}_{\mathrm{f}}$ of $|\Psi^{(n)}_{\mathrm{f}}\rangle$  can be conveniently expressed as $\mathcal{N}_{\mathrm{f}}=\sqrt{\mathrm{det}(\mathcal{M}^{(n)})}$ where $\mathrm{det}(\mathcal{M}^{(n)})$ is the determinant of the $n\times n$ matrix
\begin{equation}\label{matrixM}
\mathcal{M}^{(n)}=
\begin{pmatrix}
\langle \alpha_1 \downarrow|\alpha_1 \downarrow \rangle & \cdots  & \langle \alpha_1 \downarrow|\alpha_N \uparrow \rangle \\
\vdots &  \ddots &  \vdots \\
\langle \alpha_N \uparrow|\alpha_1 \downarrow \rangle  & \cdots  & \langle \alpha_N \uparrow|\alpha_N \uparrow \rangle
\end{pmatrix},
\end{equation}
defined in the $n$-dimensional one-particle basis $\{\ket{\alpha_1 \downarrow},\ket{\alpha_1 \uparrow},\ket{\alpha_2 \downarrow},\ket{\alpha_2 \uparrow},\ldots,\ket{\alpha_N\downarrow},\ket{\alpha_N\uparrow}\}$.

The sLOCC framework is once again realized by counting one particle in each of the nodes A and B (this entails that two particles are in each node M$_i$), also allowing for classical communication of the counting outcomes. The post-selected global state results to be
\begin{equation}
|\Psi^{(n)}_{\mathrm{f,PS}}\rangle=
|\Psi_{\mathrm{M}_1},\Psi_{\mathrm{M}_2},\ldots,\Psi_{\mathrm{M}_k}, \Psi_{\mathrm{AB}}^{-}\rangle,
\end{equation}
where $\ket{\Psi_{\mathrm{M}_i}}=\ket{\mathrm{M}_i\uparrow,\mathrm{M}_i\downarrow}$ and $\ket{\Psi_{\mathrm{AB}}^{-}}$ is a maximally entangled Bell state (see Eq.~\eqref{Bell}). The probability to obtain the state above, that is the success probability of the remote entanglement transfer process, is given by $P_{\mathrm{f}}(n)=|\langle \Psi_{\mathrm{f,PS}}^{(n)}|\Psi_{\mathrm{f}}^{(n)}\rangle|^2$. It is straightforward to show that its explicit expression as a function of the number of fermions is
\begin{equation}
P_{\mathrm{f}}(n)=\frac{1}{2^{n-1}\mathrm{det}(\mathcal{M}^{(n)})}.
\end{equation}
\section{Scalar products in the no-label formalism}\label{AppA}

For calculating the success probabilities of the proposed protocols under  different configurations, we need to compute scalar products between states of $n$ identical particles. These are obtained by  the $n$-particle probability amplitude defined in the non-standard approach (no-label particle-based formalism) here adopted \cite{LFCSciRep,compagno2018dealing}, whose general expression is
\begin{eqnarray}\label{Nampleta}
&\langle \varphi'_1,\varphi'_2,\ldots,\varphi'_n|\varphi_1,\varphi_2,\ldots,\varphi_n\rangle & \nonumber\\ & :=\sum_P\eta^P\langle \varphi'_1|\varphi_{P_1}\rangle\langle \varphi'_2|\varphi_{P_2}\rangle \ldots \langle \varphi'_n|\varphi_{P_n}\rangle, &
\end{eqnarray}
where $P=\{P_1,P_2,...,P_n\}$ in the sum runs over all the one-particle state permutations, $\eta=\pm1$ for bosons and fermions, respectively, and $\eta^P$ is 1 for bosons and 1 (-1) for even (odd) permutations for fermions.

\section{Explicit calculations for shared intermediate sites with bosons}\label{AppB}
The four-boson global state $|\Psi^{(4)}_{\mathrm{b}}\rangle$ is
\begin{equation}
|\Psi^{(4)}_{\mathrm{b}}\rangle=\dfrac{1}{5}|\mathrm{(A+M)\downarrow,(A+M)\uparrow,(M+B)\downarrow,(M+B)\uparrow\rangle}.
\end{equation}
The basis for sLOCC, corresponding to counting  two particles in the shared intermediate mode M and one particle in each mode A and B, is $\mathcal{B}_{\mathrm{b}}=\Bigl\{\dfrac{|\mathrm{A \ \sigma, M \ \tau, M \ \sigma', B \ \tau'\rangle}}{\mathcal{N}_{\tau\sigma'}}\Bigr\}$ \ ($\sigma,\tau,\sigma',\tau'=\downarrow,\uparrow$) where $\mathcal{N}_{\tau\sigma'}=\sqrt{1+\langle \tau|\sigma'\rangle}$.
The four-boson post-selected state, after projection onto $\mathcal{B}_{\mathrm{b}}$ (also allowing for classical communication), is then
\begin{align}\label{finalb}
\begin{split}
|\Psi_{\mathrm{b,PS}}^{(4)}\rangle=&\dfrac{1}{\sqrt{6}}(|\mathrm{A}\downarrow,\mathrm{M}\uparrow,\mathrm{M}\uparrow,\mathrm{B}\downarrow\rangle+|\mathrm{A}\downarrow,\mathrm{M}\uparrow,\mathrm{M}\downarrow,\mathrm{B}\uparrow\rangle\\
&+|\mathrm{A}\uparrow,\mathrm{M}\uparrow,\mathrm{M}\downarrow,\mathrm{B}\downarrow\rangle+|\mathrm{A}\uparrow,\mathrm{M}\downarrow,\mathrm{M}\downarrow,\mathrm{B}\uparrow\rangle),
\end{split}
\end{align}
which is found with probability $P_{\mathrm{b}}(4)=|\langle \Psi^{(4)}_{\mathrm{b,PS}}|\Psi^{(4)}_{\mathrm{b}}\rangle|^2=6/25$.

In the M-subspace, the Bell basis for bosons is given by the three states $|\Phi^{\pm}_{\mathrm{M}}\rangle=(|\mathrm{M}\downarrow,\mathrm{M}\downarrow\rangle\pm|\mathrm{M}\uparrow,\mathrm{M}\uparrow\rangle)/2$ and $\ket{\Psi_{\mathrm{M}}}=\ket{\mathrm{M}\uparrow,\mathrm{M}\downarrow}$. The post-selected state $|\Psi^{(4)}_{\mathrm{b,PS}}\rangle$ can be then expressed in terms of this Bell basis as
\begin{equation}\label{finalb2}
|\Psi_{\mathrm{b,PS}}^{(4)}\rangle=
\frac{\ket{\Psi_{\mathrm{M}},\Psi^{+}_{\mathrm{AB}}}
+\ket{\Phi^{+}_{\mathrm{M}},\Phi^{+}_{\mathrm{AB}}}
- \ket{\Phi^{-}_{\mathrm{M}},\Phi^{-}_{\mathrm{AB}}}}{\sqrt{3}},
\end{equation}
where in each term the particles in the distant sites A and B are in a Bell state (see Eq.~\eqref{Bell}). Therefore, each outcome of the joint Bell measurement successfully realizes the entanglement swapping over the distant nodes A and B.

The four-particle bosonic protocol can be extended, analogously to the standard ES, by a cascaded procedure \cite{2008multistage}. The scheme is again that of Fig. \ref{Fig2} with $n$ independently-prepared identical bosons and $k=N-1$ intermediate nodes ($N=n/2$ is the number of particle pairs).

The initially prepared $n$-boson state is
\begin{equation}\label{initPsiboson}
|\Psi^{(n)}_{\mathrm{b}}\rangle=\frac{1}{\mathcal{N}_{b}}|\alpha_1\downarrow,\alpha_1\uparrow, \alpha_2 \downarrow,\alpha_2 \uparrow, \ldots, \alpha_N \downarrow,\alpha_N\uparrow\rangle,
\end{equation}
where the normalization constant $\mathcal{N}_{b}=\sqrt{\mathrm{perm}(\mathcal{M}^{(n)})}$, with $\mathrm{perm}(\mathcal{M}^{(n)})$ being the permanent of the matrix of Eq.~\eqref{matrixM}.

\begin{figure}[!t]
\begin{center}
\includegraphics[scale=0.47]{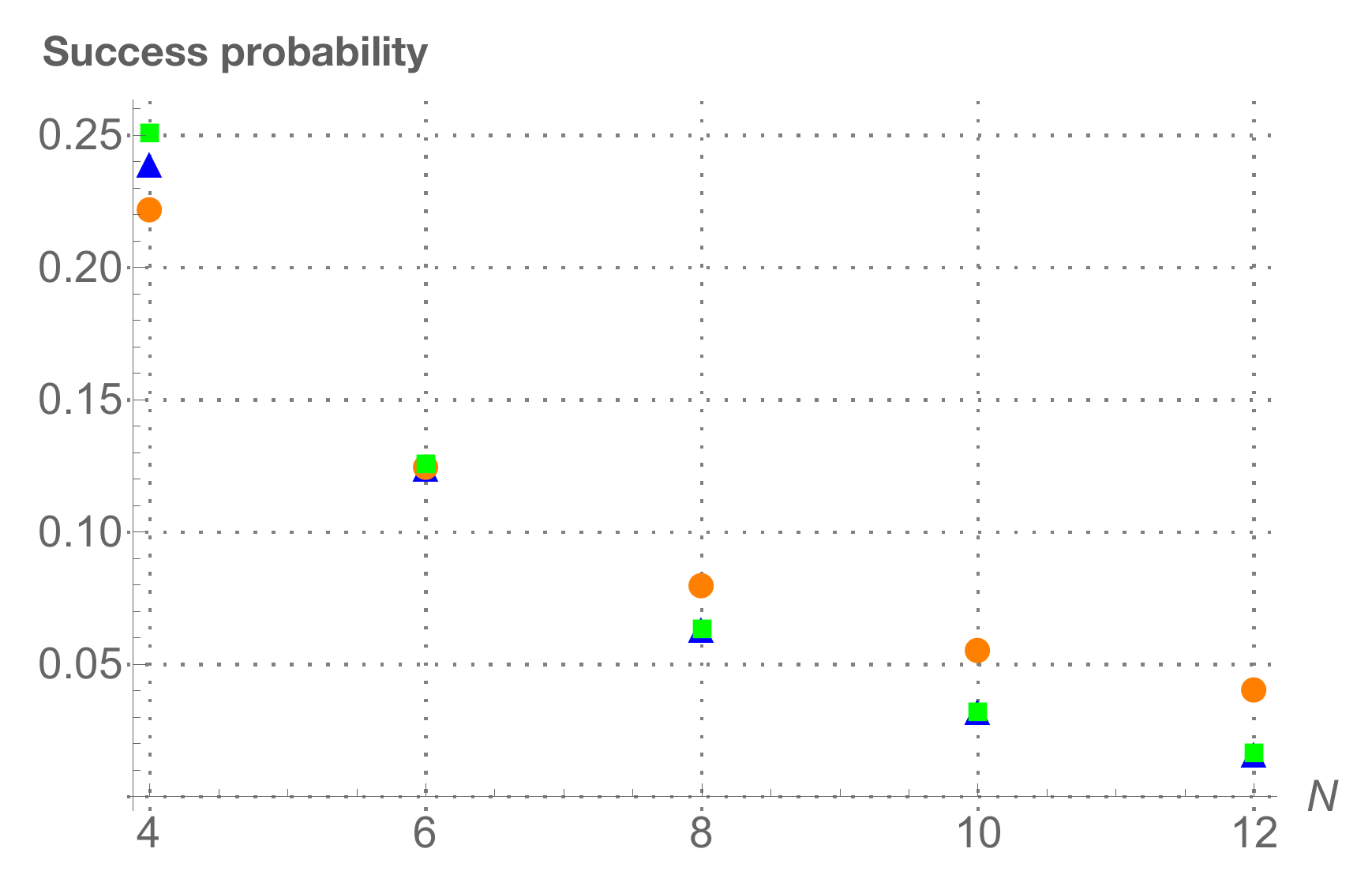}
\caption{Success probability to implement multiple entanglement transfer as a function of the number of particles $n$, for separated
nodes with either bosons or fermions ($P(n)$, green squares) and for shared intermediate nodes with fermions ($P_{\mathrm{f}}(n)$, orange points) and with bosons ($P_{\mathrm{b}}(n)$, blue triangles).}
\label{Fig4}
\end{center}
\end{figure}

By counting two particles in each intermediate node and one in each of the distant nodes A and B, also allowing for classical communication of the counting outcomes, one gets the post-selected state $\ket{\Psi_{\mathrm{b,PS}}^{(n)}}$.  Bell measurements are then performed step by step on each intermediate node $\mathrm{M}_i$ ($i=1,...,k$) to transfer entanglement over A and B. The type of the final Bell state transferred over A and B will depend on the consecutive outcomes of the cascaded Bell measurements. The success probability of the protocol is obtained by $P_{\mathrm{b}}(n)=|\langle \Psi_{\mathrm{b,PS}}^{(n)}|\Psi_{\mathrm{b}}^{(n)}\rangle|^2$. Its explicit expression as a function of the number of bosons, in the cases of Fig.~\ref{Fig4}, is
\begin{equation}
P_{\mathrm{b}}(n)=\frac{3^{\frac{n}{2}-1}}{2^{n-1}\mathrm{perm}(\mathcal{M}^{(n)})}.
\end{equation}

In order to be more explicit concerning the cascaded procedure leading to the Bell states over A and B, we treat the case with $n=6$ bosons ($N=3$ pairs) and two shared intermediate nodes M$_1$, M$_2$.
From the initially prepared state $|\Psi^{(6)}_{\mathrm{b}}\rangle$, easily obtained from Eq.~\eqref{initPsiboson}, the sLOCC framework counting two particles in the intermediate nodes and one particle in each of the far nodes A and B, including classical communication, leads to the post-selected state
\begin{eqnarray}\label{PS6boson}
|\Psi_{\mathrm{b,PS}}^{(6)}\rangle &=&
\frac{\sqrt{2}}{3}(\ket{\Psi_{\mathrm{M_1}},\Psi^{+}_{\mathrm{AM_2}}}
+\ket{\Phi^{+}_{\mathrm{M_1}},\Phi^{+}_{\mathrm{AM_2}}}
\nonumber\\
&&-\ket{\Phi^{-}_{\mathrm{M_1}},\Phi^{-}_{\mathrm{AM_2}}})
\wedge \ket{\Psi^{+}_{\mathrm{M_2 B}}},
\end{eqnarray}
where the relevant Bell states are analogous to those given after Eq.~\eqref{finalb} and in Eq.~\eqref{Bell}.
A first Bell measurement has to be performed on the intermediate node M$_1$ in order to entangle bosons in A and M$_2$. Any outcome is good for continuing the protocol. Let us suppose, without loss of generality, that the result of this first Bell measurement is $\ket{\Phi^{-}_{\mathrm{M_1}}}$. From $|\Psi_{\mathrm{b,PS}}^{(6)}\rangle$ above, one sees that the remaining four bosons are left into the state $-\ket{\Phi^{-}_{\mathrm{AM_2}},\Psi^{+}_{\mathrm{M_2 B}}}$. This state, suitably normalized and expressed in terms of the Bell basis in the node M$_2$, assumes the form
\begin{equation}\label{finalb6}
\ket{\Psi^{(4)}}_\mathrm{AM_2B}=
\frac{\ket{\Phi^{+}_{\mathrm{M}_2},\Psi^{-}_{\mathrm{AB}}}
+\ket{\Phi^{-}_{\mathrm{M}_2},\Psi^{+}_{\mathrm{AB}}}
+\ket{\Psi_{\mathrm{M}_2},\Phi^{-}_{\mathrm{AB}}}}{\sqrt{3}}.
\end{equation}
It is now clear that a second Bell measurement on the intermediate node M$_2$ has the final effect to transfer a Bell state over the far nodes A and B. Thus, for six bosons, two cascaded Bell measurements realize the desired entanglement swapping protocol. This procedure can be continued analogously for successive steps with more particles.

\section{Explicit calculations for separated intermediate sites}\label{AppSeparated}

Here, we report the calculations regarding the scheme above, depicted in Fig. \ref{Fig3} .
The prepared state $|\Psi^{(4)}\rangle$ can be written as the superposition of 16 terms
\begin{align}\label{Psiexpansion}
\begin{split}
|\Psi^{(4)}\rangle=&\dfrac{1}{4}(\mathrm{|A\downarrow,A\uparrow,D\downarrow,D\uparrow\rangle+|A\downarrow,A\uparrow,D\downarrow,B\uparrow\rangle}\\
&+\mathrm{|A\downarrow,A\uparrow,B\downarrow,D\uparrow\rangle+|A\downarrow,A\uparrow,B\downarrow,B\uparrow\rangle}\\
&+\mathrm{|A\downarrow,C\uparrow,D\downarrow,D\uparrow\rangle+|A\downarrow,C\uparrow,D\downarrow,B\uparrow\rangle}\\
&+\mathrm{|A\downarrow,C\uparrow,B\downarrow,D\uparrow\rangle+|A\downarrow,C\uparrow,B\downarrow,B\uparrow\rangle}\\
&+\mathrm{|C\downarrow,A\uparrow,D\downarrow,D\uparrow\rangle+|C\downarrow,A\uparrow,D\downarrow,B\uparrow\rangle}\\
&+\mathrm{|C\downarrow,A\uparrow,B\downarrow,D\uparrow\rangle+|C\downarrow,A\uparrow,B\downarrow,B\uparrow\rangle}\\
&+\mathrm{|C\downarrow,C\uparrow,D\downarrow,D\uparrow\rangle+|C\downarrow,C\uparrow,D\downarrow,B\uparrow\rangle}\\
&+\mathrm{|C\downarrow,C\uparrow,B\downarrow,D\uparrow\rangle+|C\downarrow,C\uparrow,B\downarrow,B\uparrow\rangle}
).
\end{split}
\end{align}

In the linear combination of Eq.~\eqref{Psiexpansion}, there are contributions in which two particles occupy the same site. We perform sLOCC in the form of a post-selection counting a single particle in each site A, B and classically communicating their outcomes to each other (notice that a single particle in A entails one particle in C and a single particle in B implies one particle in D). This post-selection corresponds to project the global four-particle state $|\Psi^{(4)}\rangle$ onto the subspace spanned by the spatially localized basis $\mathcal{B}=\{|\mathrm{A \ \sigma,C \ \tau,D \ \sigma',B \ \tau'\rangle}\}$, by the projector $\hat{\Pi}_{\mathrm{ACDB}}=\sum_{\sigma,\tau,\sigma',\tau'=\downarrow,\uparrow}|\mathrm{A \ \sigma,C \ \tau,D \ \sigma',B \ \tau'}\rangle \langle \mathrm{A \ \sigma,C \ \tau,D \ \sigma',B \ \tau'|}$. The post-selected (projected) state is thus obtained as
\begin{equation}
|\Psi^{(4)}_{\mathrm{PS}}\rangle=\hat{\Pi}_{\mathrm{ACDB}}|\Psi^{(4)}\rangle/\mathcal{N},
\end{equation}
where $\mathcal{N}=\sqrt{\langle \Psi^{(4)}|\hat{\Pi}_{\mathrm{ACDB}}|\Psi^{(4)}\rangle}=1/2$.
Its explicit expression is
\begin{align}\label{PostABCD}
\begin{split}
|\Psi^{(4)}_{\mathrm{PS}}\rangle=&\dfrac{1}{\sqrt{2}}(|\mathrm{A\downarrow,C\uparrow}\rangle +\eta|\mathrm{A\uparrow,C\downarrow\rangle})\\
&\wedge \dfrac{1}{\sqrt{2}}(|\mathrm{D\downarrow,B\uparrow\rangle+\eta|D\uparrow,B\downarrow}\rangle),
\end{split}
\end{align}
that is $|\Psi^{(4)}_{\mathrm{PS}}\rangle=|\Psi_{\mathrm{AC}},\Psi_{\mathrm{DB}}\rangle$ (see Eqs.~\eqref{PsiPStensor} and \eqref{BellStates} above). In Eq.~\eqref{PostABCD} we have used the \textit{wedge} product $\wedge$ that, in this case of separated sites under sLOCC, coincides with the standard tensor product \cite{compagno2018dealing}. This state is obtained with probability $P(4)=|\langle \Psi_{\mathrm{PS}}^{(4)}|\Psi^{(4)}\rangle|^2=1/4$.
Notice that in this sLOCC framework, the prepared state $|\Psi^{(4)}\rangle$ can be written as $|\Psi^{(4)}\rangle=\ket{\alpha \downarrow,\alpha \uparrow}\wedge\ket{\beta \downarrow, \beta \uparrow}$, from which one then obtains particle entanglement between (A, C) and (D, B), as evidenced in Eq.~\eqref{PostABCD}. This is linked to the concept of indistinguishability as a resource by sLOCC introduced in Ref. \cite{LoFrancoPRL}.

Since the sites (A, C) and (D, B) are separated, the identical particles can be distinguished by their spatial location. Once got the post-selected state $|\Psi^{(4)}_{\mathrm{PS}}\rangle$, the entanglement swapping proceeds following the lines of the standard protocol for distinguishable particles \citep{1zuk1993}. A Bell measurement is therefore performed on the  intermediate nodes (C, D) to obtain entanglement over the far nodes A and B. In fact, one can write
\begin{align}
\begin{split}
|\Psi^{(4)}_{\mathrm{PS}}\rangle&=\dfrac{1}{2}[|\Psi^+_{\mathrm{CD}},\Psi^+_{\mathrm{AB}}\rangle-|\Psi^-_{\mathrm{CD}},\Psi^-_{\mathrm{AB}}\rangle\\
&+\eta|\Phi^+_{\mathrm{CD}},\Phi^+_{\mathrm{AB}}\rangle-\eta|\Phi^-_{\mathrm{CD}},\Phi^-_{\mathrm{AB}}\rangle],
\end{split}
\end{align}
where
\begin{align}\label{Bell}
\begin{split}
&|\Psi^{\pm}_{\mathrm{IJ}}\rangle=\dfrac{1}{\sqrt{2}}(\mathrm{|I\downarrow,J\uparrow\rangle \pm |I\uparrow,J\downarrow}\rangle),\\
&|\Phi^{\pm}_{\mathrm{IJ}}\rangle=\dfrac{1}{\sqrt{2}}(\mathrm{|I\downarrow,J\downarrow\rangle \pm |I\uparrow,J \uparrow\rangle}),
\end{split}
\end{align}
with IJ$=$AB, CD. The result of the Bell measurement does not depend on the particle statistics, as expected from the fact that the post-selected state describes identical particles in separated spatial regions under sLOCC.

The previous protocol can be straightforwardly extended to multiple entanglement swapping (with $n = 2N$ independently prepared particles, being $N$ the
number of involved particle pairs), in analogy to the case of distinguishable particles \cite{2008multistage}, by a cascaded procedure with a success probability $P(n)=1/2^{\frac{n}{2}}$.

\section{Probabilities of success}

We finally compare the efficiency of the protocol for the various cases treated above, given the  prepared state before post-selection. We have already seen that, in the case of separated nodes, the success probability for both bosons and fermions is $P(n)=1/2^{\frac{n}{2}}$ (see green squares in Fig.~\ref{Fig4}).
In the case of shared intermediate nodes, the success probabilities $P_{\mathrm{f}}(n)$ for fermions and $P_{\mathrm{b}}(n)$ for bosons decrease as a function of the particle number similarly to $P(n)$, as displayed in Fig.~\ref{Fig4} (orange points and blue triangles, respectively).
From the experimental viewpoint, one has to take into account that the requirement of Bell measurements further hinders the protocol efficiency \citep{yang2006experimental,BSA,lee2013bell,cQED,cQED3,PhysRevA.84.042331,PhysRevA.92.042314,PhysRevA.82.032318}. Therefore, the fermionic process results in being not only qualitatively different, but also more advantageous from a practical viewpoint than the other procedures that necessarily require Bell measurements.

\begin{thebibliography}{60}
\expandafter\ifx\csname natexlab\endcsname\relax\def\natexlab#1{#1}\fi
\expandafter\ifx\csname bibnamefont\endcsname\relax
  \def\bibnamefont#1{#1}\fi
\expandafter\ifx\csname bibfnamefont\endcsname\relax
  \def\bibfnamefont#1{#1}\fi
\expandafter\ifx\csname citenamefont\endcsname\relax
  \def\citenamefont#1{#1}\fi
\expandafter\ifx\csname url\endcsname\relax
  \def\url#1{\texttt{#1}}\fi
\expandafter\ifx\csname urlprefix\endcsname\relax\def\urlprefix{URL }\fi
\providecommand{\bibinfo}[2]{#2}
\providecommand{\eprint}[2][]{\url{#2}}

\bibitem[{\citenamefont{Gisin and Thew}(2007)}]{quantumRepeaters}
\bibinfo{author}{\bibfnamefont{N.}~\bibnamefont{Gisin}} \bibnamefont{and}
  \bibinfo{author}{\bibfnamefont{R.}~\bibnamefont{Thew}},
  \bibinfo{journal}{Nat. Photon.} \textbf{\bibinfo{volume}{1}},
  \bibinfo{pages}{165} (\bibinfo{year}{2007}).

\bibitem[{\citenamefont{Castelvecchi}(2018)}]{castelvecchiQuantInternet}
\bibinfo{author}{\bibfnamefont{D.}~\bibnamefont{Castelvecchi}},
  \bibinfo{journal}{Nature} \textbf{\bibinfo{volume}{554}},
  \bibinfo{pages}{289} (\bibinfo{year}{2018}).

\bibitem[{\citenamefont{Pirandola and
  Braunstein}(2016)}]{pirandolaQuantInternet}
\bibinfo{author}{\bibfnamefont{S.}~\bibnamefont{Pirandola}} \bibnamefont{and}
  \bibinfo{author}{\bibfnamefont{S.~L.} \bibnamefont{Braunstein}},
  \bibinfo{journal}{Nature} \textbf{\bibinfo{volume}{532}},
  \bibinfo{pages}{169} (\bibinfo{year}{2016}).

\bibitem[{\citenamefont{Humphreys et~al.}(2018)\citenamefont{Humphreys, Kalb,
  Morits, Schouten, Vermeulen, Twitchen, Markham, and
  Hanson}}]{humphreys2018deterministic}
\bibinfo{author}{\bibfnamefont{P.~C.} \bibnamefont{Humphreys}},
  \bibinfo{author}{\bibfnamefont{N.}~\bibnamefont{Kalb}},
  \bibinfo{author}{\bibfnamefont{J.~P.~J.} \bibnamefont{Morits}},
  \bibinfo{author}{\bibfnamefont{R.~N.} \bibnamefont{Schouten}},
  \bibinfo{author}{\bibfnamefont{R.~F.~L.} \bibnamefont{Vermeulen}},
  \bibinfo{author}{\bibfnamefont{D.~J.} \bibnamefont{Twitchen}},
  \bibinfo{author}{\bibfnamefont{M.}~\bibnamefont{Markham}}, \bibnamefont{and}
  \bibinfo{author}{\bibfnamefont{R.}~\bibnamefont{Hanson}},
  \bibinfo{journal}{Nature} \textbf{\bibinfo{volume}{558}},
  \bibinfo{pages}{268} (\bibinfo{year}{2018}).

\bibitem[{\citenamefont{Omar}(2005)}]{omarIJQI}
\bibinfo{author}{\bibfnamefont{Y.}~\bibnamefont{Omar}}, \bibinfo{journal}{Int.
  J. Quantum Inform.} \textbf{\bibinfo{volume}{3}}, \bibinfo{pages}{201}
  (\bibinfo{year}{2005}).

\bibitem[{\citenamefont{Omar et~al.}(2002)\citenamefont{Omar,
  Paunkovi\ifmmode~\acute{c}\else \'{c}\fi{}, Bose, and
  Vedral}}]{PhysRevA.65.062305}
\bibinfo{author}{\bibfnamefont{Y.}~\bibnamefont{Omar}},
  \bibinfo{author}{\bibfnamefont{N.}~\bibnamefont{Paunkovi\ifmmode~\acute{c}\else
  \'{c}\fi{}}}, \bibinfo{author}{\bibfnamefont{S.}~\bibnamefont{Bose}},
  \bibnamefont{and} \bibinfo{author}{\bibfnamefont{V.}~\bibnamefont{Vedral}},
  \bibinfo{journal}{Phys. Rev. A} \textbf{\bibinfo{volume}{65}},
  \bibinfo{pages}{062305} (\bibinfo{year}{2002}).

\bibitem[{\citenamefont{Paunkovi\ifmmode~\acute{c}\else \'{c}\fi{}
  et~al.}(2002)\citenamefont{Paunkovi\ifmmode~\acute{c}\else \'{c}\fi{}, Omar,
  Bose, and Vedral}}]{PhysRevLett.88.187903}
\bibinfo{author}{\bibfnamefont{N.}~\bibnamefont{Paunkovi\ifmmode~\acute{c}\else
  \'{c}\fi{}}}, \bibinfo{author}{\bibfnamefont{Y.}~\bibnamefont{Omar}},
  \bibinfo{author}{\bibfnamefont{S.}~\bibnamefont{Bose}}, \bibnamefont{and}
  \bibinfo{author}{\bibfnamefont{V.}~\bibnamefont{Vedral}},
  \bibinfo{journal}{Phys. Rev. Lett.} \textbf{\bibinfo{volume}{88}},
  \bibinfo{pages}{187903} (\bibinfo{year}{2002}).

\bibitem[{\citenamefont{Bose et~al.}(2003)\citenamefont{Bose, Ekert, Omar,
  Paunkovi\ifmmode~\acute{c}\else \'{c}\fi{}, and Vedral}}]{PhysRevA.68.052309}
\bibinfo{author}{\bibfnamefont{S.}~\bibnamefont{Bose}},
  \bibinfo{author}{\bibfnamefont{A.}~\bibnamefont{Ekert}},
  \bibinfo{author}{\bibfnamefont{Y.}~\bibnamefont{Omar}},
  \bibinfo{author}{\bibfnamefont{N.}~\bibnamefont{Paunkovi\ifmmode~\acute{c}\else
  \'{c}\fi{}}}, \bibnamefont{and}
  \bibinfo{author}{\bibfnamefont{V.}~\bibnamefont{Vedral}},
  \bibinfo{journal}{Phys. Rev. A} \textbf{\bibinfo{volume}{68}},
  \bibinfo{pages}{052309} (\bibinfo{year}{2003}).

\bibitem[{\citenamefont{Bose and Home}(2002)}]{bose2002indisting}
\bibinfo{author}{\bibfnamefont{S.}~\bibnamefont{Bose}} \bibnamefont{and}
  \bibinfo{author}{\bibfnamefont{D.}~\bibnamefont{Home}},
  \bibinfo{journal}{Phys. Rev. Lett.} \textbf{\bibinfo{volume}{88}},
  \bibinfo{pages}{050401} (\bibinfo{year}{2002}).

\bibitem[{\citenamefont{{Lo Franco} and Compagno}(2016)}]{LFCSciRep}
\bibinfo{author}{\bibfnamefont{R.}~\bibnamefont{{Lo Franco}}} \bibnamefont{and}
  \bibinfo{author}{\bibfnamefont{G.}~\bibnamefont{Compagno}},
  \bibinfo{journal}{Sci. Rep.} \textbf{\bibinfo{volume}{6}},
  \bibinfo{pages}{20603} (\bibinfo{year}{2016}).

\bibitem[{\citenamefont{Sciara et~al.}(2017)\citenamefont{Sciara, {Lo Franco},
  and Compagno}}]{sciaraSchmidt}
\bibinfo{author}{\bibfnamefont{S.}~\bibnamefont{Sciara}},
  \bibinfo{author}{\bibfnamefont{R.}~\bibnamefont{{Lo Franco}}},
  \bibnamefont{and} \bibinfo{author}{\bibfnamefont{G.}~\bibnamefont{Compagno}},
  \bibinfo{journal}{Sci. Rep.} \textbf{\bibinfo{volume}{7}},
  \bibinfo{pages}{44675} (\bibinfo{year}{2017}).

\bibitem[{\citenamefont{Bellomo et~al.}(2017)\citenamefont{Bellomo, {Lo
  Franco}, and Compagno}}]{bellomo2017}
\bibinfo{author}{\bibfnamefont{B.}~\bibnamefont{Bellomo}},
  \bibinfo{author}{\bibfnamefont{R.}~\bibnamefont{{Lo Franco}}},
  \bibnamefont{and} \bibinfo{author}{\bibfnamefont{G.}~\bibnamefont{Compagno}},
  \bibinfo{journal}{Phys. Rev. A} \textbf{\bibinfo{volume}{96}},
  \bibinfo{pages}{022319} (\bibinfo{year}{2017}).

\bibitem[{\citenamefont{Compagno et~al.}(2018)\citenamefont{Compagno,
  Castellini, and {Lo Franco}}}]{compagno2018dealing}
\bibinfo{author}{\bibfnamefont{G.}~\bibnamefont{Compagno}},
  \bibinfo{author}{\bibfnamefont{A.}~\bibnamefont{Castellini}},
  \bibnamefont{and} \bibinfo{author}{\bibfnamefont{R.}~\bibnamefont{{Lo
  Franco}}}, \bibinfo{journal}{Phil. Trans. R. Soc. A}
  \textbf{\bibinfo{volume}{376}}, \bibinfo{pages}{20170317}
  (\bibinfo{year}{2018}).

\bibitem[{\citenamefont{{Lo Franco} and Compagno}(2018)}]{LoFrancoPRL}
\bibinfo{author}{\bibfnamefont{R.}~\bibnamefont{{Lo Franco}}} \bibnamefont{and}
  \bibinfo{author}{\bibfnamefont{G.}~\bibnamefont{Compagno}},
  \bibinfo{journal}{Phys. Rev. Lett.} \textbf{\bibinfo{volume}{120}},
  \bibinfo{pages}{240403} (\bibinfo{year}{2018}).

\bibitem[{\citenamefont{Benatti et~al.}(2014)\citenamefont{Benatti, Alipour,
  and Rezakhani}}]{benatti2014NJP}
\bibinfo{author}{\bibfnamefont{F.}~\bibnamefont{Benatti}},
  \bibinfo{author}{\bibfnamefont{S.}~\bibnamefont{Alipour}}, \bibnamefont{and}
  \bibinfo{author}{\bibfnamefont{A.~T.} \bibnamefont{Rezakhani}},
  \bibinfo{journal}{New J. Phys.} \textbf{\bibinfo{volume}{16}},
  \bibinfo{pages}{015023} (\bibinfo{year}{2014}).


\bibitem[{\citenamefont{Braun et~al.}(2018)\citenamefont{Braun, Adesso,
  Benatti, Floreanini, Marzolino, Mitchell, and Pirandola}}]{QEMreview}
\bibinfo{author}{\bibfnamefont{D.}~\bibnamefont{Braun}},
  \bibinfo{author}{\bibfnamefont{G.}~\bibnamefont{Adesso}},
  \bibinfo{author}{\bibfnamefont{F.}~\bibnamefont{Benatti}},
  \bibinfo{author}{\bibfnamefont{R.}~\bibnamefont{Floreanini}},
  \bibinfo{author}{\bibfnamefont{U.}~\bibnamefont{Marzolino}},
  \bibinfo{author}{\bibfnamefont{M.~W.} \bibnamefont{Mitchell}},
  \bibnamefont{and}
  \bibinfo{author}{\bibfnamefont{S.}~\bibnamefont{Pirandola}},
  \bibinfo{journal}{Rev. Mod. Phys.} \textbf{\bibinfo{volume}{90}},
  \bibinfo{pages}{035006} (\bibinfo{year}{2018}).

\bibitem[{\citenamefont{Scherer et~al.}(2011)\citenamefont{Scherer, Sanders,
  and Tittel}}]{quantcomm}
\bibinfo{author}{\bibfnamefont{A.}~\bibnamefont{Scherer}},
  \bibinfo{author}{\bibfnamefont{B.~C.} \bibnamefont{Sanders}},
  \bibnamefont{and} \bibinfo{author}{\bibfnamefont{W.}~\bibnamefont{Tittel}},
  \bibinfo{journal}{Opt. Express} \textbf{\bibinfo{volume}{19}},
  \bibinfo{pages}{3004} (\bibinfo{year}{2011}).

\bibitem[{\citenamefont{Ye}(2015)}]{quantcomm2}
\bibinfo{author}{\bibfnamefont{T.-Y.} \bibnamefont{Ye}},
  \bibinfo{journal}{Quant. Inform. Process.} \textbf{\bibinfo{volume}{14}},
  \bibinfo{pages}{1469} (\bibinfo{year}{2015}).

\bibitem[{\citenamefont{Naseri}(2016)}]{quantcomm3}
\bibinfo{author}{\bibfnamefont{M.}~\bibnamefont{Naseri}},
  \bibinfo{journal}{Int. J. Theor. Phys.} \textbf{\bibinfo{volume}{55}},
  \bibinfo{pages}{2428} (\bibinfo{year}{2016}).

\bibitem[{\citenamefont{Sun et~al.}(2017{\natexlab{a}})\citenamefont{Sun, Mao,
  Jiang, Zhao, Chen, Zhang, Zhang, Jiang, Chen, You
  et~al.}}]{sun2017entanglement}
\bibinfo{author}{\bibfnamefont{Q.-C.} \bibnamefont{Sun}},
  \bibinfo{author}{\bibfnamefont{Y.-L.} \bibnamefont{Mao}},
  \bibinfo{author}{\bibfnamefont{Y.-F.} \bibnamefont{Jiang}},
  \bibinfo{author}{\bibfnamefont{Q.}~\bibnamefont{Zhao}},
  \bibinfo{author}{\bibfnamefont{S.-J.} \bibnamefont{Chen}},
  \bibinfo{author}{\bibfnamefont{W.}~\bibnamefont{Zhang}},
  \bibinfo{author}{\bibfnamefont{W.-J.} \bibnamefont{Zhang}},
  \bibinfo{author}{\bibfnamefont{X.}~\bibnamefont{Jiang}},
  \bibinfo{author}{\bibfnamefont{T.-Y.} \bibnamefont{Chen}},
  \bibinfo{author}{\bibfnamefont{L.-X.} \bibnamefont{You}},
  \bibnamefont{et~al.}, \bibinfo{journal}{Phys. Rev. A}
  \textbf{\bibinfo{volume}{95}}, \bibinfo{pages}{032306}
  (\bibinfo{year}{2017}{\natexlab{a}}).

\bibitem[{\citenamefont{Pan et~al.}(1998)\citenamefont{Pan, Bouwmeester,
  Weinfurter, and Zeilinger}}]{pan1998experimental}
\bibinfo{author}{\bibfnamefont{J.-W.} \bibnamefont{Pan}},
  \bibinfo{author}{\bibfnamefont{D.}~\bibnamefont{Bouwmeester}},
  \bibinfo{author}{\bibfnamefont{H.}~\bibnamefont{Weinfurter}},
  \bibnamefont{and}
  \bibinfo{author}{\bibfnamefont{A.}~\bibnamefont{Zeilinger}},
  \bibinfo{journal}{Phys. Rev. Lett.} \textbf{\bibinfo{volume}{80}},
  \bibinfo{pages}{3891} (\bibinfo{year}{1998}).

\bibitem[{\citenamefont{de~Riedmatten et~al.}(2005)\citenamefont{de~Riedmatten,
  Marcikic, van Houwelingen, Tittel, Zbinden, and Gisin}}]{experiment}
\bibinfo{author}{\bibfnamefont{H.}~\bibnamefont{de~Riedmatten}},
  \bibinfo{author}{\bibfnamefont{I.}~\bibnamefont{Marcikic}},
  \bibinfo{author}{\bibfnamefont{J.~A.~W.} \bibnamefont{van Houwelingen}},
  \bibinfo{author}{\bibfnamefont{W.}~\bibnamefont{Tittel}},
  \bibinfo{author}{\bibfnamefont{H.}~\bibnamefont{Zbinden}}, \bibnamefont{and}
  \bibinfo{author}{\bibfnamefont{N.}~\bibnamefont{Gisin}},
  \bibinfo{journal}{Phys. Rev. A} \textbf{\bibinfo{volume}{71}},
  \bibinfo{pages}{050302} (\bibinfo{year}{2005}).

\bibitem[{\citenamefont{Yang et~al.}(2006)\citenamefont{Yang, Zhang, Chen, Lu,
  Yin, Pan, Wei, Tian, and Zhang}}]{yang2006experimental}
\bibinfo{author}{\bibfnamefont{T.}~\bibnamefont{Yang}},
  \bibinfo{author}{\bibfnamefont{Q.}~\bibnamefont{Zhang}},
  \bibinfo{author}{\bibfnamefont{T.-Y.} \bibnamefont{Chen}},
  \bibinfo{author}{\bibfnamefont{S.}~\bibnamefont{Lu}},
  \bibinfo{author}{\bibfnamefont{J.}~\bibnamefont{Yin}},
  \bibinfo{author}{\bibfnamefont{J.-W.} \bibnamefont{Pan}},
  \bibinfo{author}{\bibfnamefont{Z.-Y.} \bibnamefont{Wei}},
  \bibinfo{author}{\bibfnamefont{J.-R.} \bibnamefont{Tian}}, \bibnamefont{and}
  \bibinfo{author}{\bibfnamefont{J.}~\bibnamefont{Zhang}},
  \bibinfo{journal}{Phys. Rev. Lett.} \textbf{\bibinfo{volume}{96}},
  \bibinfo{pages}{110501} (\bibinfo{year}{2006}).

\bibitem[{\citenamefont{Kaltenbaek et~al.}(2009)\citenamefont{Kaltenbaek,
  Prevedel, Aspelmeyer, and Zeilinger}}]{exp3}
\bibinfo{author}{\bibfnamefont{R.}~\bibnamefont{Kaltenbaek}},
  \bibinfo{author}{\bibfnamefont{R.}~\bibnamefont{Prevedel}},
  \bibinfo{author}{\bibfnamefont{M.}~\bibnamefont{Aspelmeyer}},
  \bibnamefont{and}
  \bibinfo{author}{\bibfnamefont{A.}~\bibnamefont{Zeilinger}},
  \bibinfo{journal}{Phys. Rev. A} \textbf{\bibinfo{volume}{79}},
  \bibinfo{pages}{040302} (\bibinfo{year}{2009}).

\bibitem[{\citenamefont{Megidish et~al.}(2013)\citenamefont{Megidish, Halevy,
  Shacham, Dvir, Dovrat, and Eisenberg}}]{megidish2013}
\bibinfo{author}{\bibfnamefont{E.}~\bibnamefont{Megidish}},
  \bibinfo{author}{\bibfnamefont{A.}~\bibnamefont{Halevy}},
  \bibinfo{author}{\bibfnamefont{T.}~\bibnamefont{Shacham}},
  \bibinfo{author}{\bibfnamefont{T.}~\bibnamefont{Dvir}},
  \bibinfo{author}{\bibfnamefont{L.}~\bibnamefont{Dovrat}}, \bibnamefont{and}
  \bibinfo{author}{\bibfnamefont{H.}~\bibnamefont{Eisenberg}},
  \bibinfo{journal}{Phys. Rev. Lett.} \textbf{\bibinfo{volume}{110}},
  \bibinfo{pages}{210403} (\bibinfo{year}{2013}).

\bibitem[{\citenamefont{Sun et~al.}(2017{\natexlab{b}})\citenamefont{Sun,
  Jiang, Mao, You, Zhang, Zhang, Jiang, Chen, Li, Huang et~al.}}]{2017timebin}
\bibinfo{author}{\bibfnamefont{Q.-C.} \bibnamefont{Sun}},
  \bibinfo{author}{\bibfnamefont{Y.-F.} \bibnamefont{Jiang}},
  \bibinfo{author}{\bibfnamefont{Y.-L.} \bibnamefont{Mao}},
  \bibinfo{author}{\bibfnamefont{L.-X.} \bibnamefont{You}},
  \bibinfo{author}{\bibfnamefont{W.}~\bibnamefont{Zhang}},
  \bibinfo{author}{\bibfnamefont{W.-J.} \bibnamefont{Zhang}},
  \bibinfo{author}{\bibfnamefont{X.}~\bibnamefont{Jiang}},
  \bibinfo{author}{\bibfnamefont{T.-Y.} \bibnamefont{Chen}},
  \bibinfo{author}{\bibfnamefont{H.}~\bibnamefont{Li}},
  \bibinfo{author}{\bibfnamefont{Y.-D.} \bibnamefont{Huang}},
  \bibnamefont{et~al.}, \bibinfo{journal}{Optica} \textbf{\bibinfo{volume}{4}},
  \bibinfo{pages}{1214} (\bibinfo{year}{2017}{\natexlab{b}}).

\bibitem[{\citenamefont{Jacobs et~al.}(2002)\citenamefont{Jacobs, Pittman, and
  Franson}}]{quantumRelay1}
\bibinfo{author}{\bibfnamefont{B.}~\bibnamefont{Jacobs}},
  \bibinfo{author}{\bibfnamefont{T.}~\bibnamefont{Pittman}}, \bibnamefont{and}
  \bibinfo{author}{\bibfnamefont{J.}~\bibnamefont{Franson}},
  \bibinfo{journal}{Phys. Rev. A} \textbf{\bibinfo{volume}{66}},
  \bibinfo{pages}{052307} (\bibinfo{year}{2002}).

\bibitem[{\citenamefont{Hu and Rarity}(2011)}]{BSA}
\bibinfo{author}{\bibfnamefont{C.}~\bibnamefont{Hu}} \bibnamefont{and}
  \bibinfo{author}{\bibfnamefont{J.}~\bibnamefont{Rarity}},
  \bibinfo{journal}{Phys. Rev. B} \textbf{\bibinfo{volume}{83}},
  \bibinfo{pages}{115303} (\bibinfo{year}{2011}).

\bibitem[{\citenamefont{Zukowski et~al.}(1993)\citenamefont{Zukowski,
  Zeilinger, Horne, and Ekert}}]{1zuk1993}
\bibinfo{author}{\bibfnamefont{M.}~\bibnamefont{Zukowski}},
  \bibinfo{author}{\bibfnamefont{A.}~\bibnamefont{Zeilinger}},
  \bibinfo{author}{\bibfnamefont{M.~A.} \bibnamefont{Horne}}, \bibnamefont{and}
  \bibinfo{author}{\bibfnamefont{A.~K.} \bibnamefont{Ekert}},
  \bibinfo{journal}{Phys. Rev. Lett.} \textbf{\bibinfo{volume}{71}},
  \bibinfo{pages}{4287} (\bibinfo{year}{1993}).

\bibitem[{\citenamefont{Jennewein et~al.}(2001)\citenamefont{Jennewein, Weihs,
  and Pan}}]{nonloc}
\bibinfo{author}{\bibfnamefont{T.}~\bibnamefont{Jennewein}},
  \bibinfo{author}{\bibfnamefont{G.}~\bibnamefont{Weihs}}, \bibnamefont{and}
  \bibinfo{author}{\bibfnamefont{A.}~\bibnamefont{Pan},
  \bibfnamefont{J.-W.and~Zeilinger}}, \bibinfo{journal}{Phys. Rev. Lett.}
  \textbf{\bibinfo{volume}{88}}, \bibinfo{pages}{017903}
  (\bibinfo{year}{2001}).

\bibitem[{\citenamefont{Branciard et~al.}(2010)\citenamefont{Branciard, Gisin,
  and Pironio}}]{nonlocality}
\bibinfo{author}{\bibfnamefont{C.}~\bibnamefont{Branciard}},
  \bibinfo{author}{\bibfnamefont{N.}~\bibnamefont{Gisin}}, \bibnamefont{and}
  \bibinfo{author}{\bibfnamefont{S.}~\bibnamefont{Pironio}},
  \bibinfo{journal}{Phys. Rev. Lett.} \textbf{\bibinfo{volume}{104}},
  \bibinfo{pages}{170401} (\bibinfo{year}{2010}).

\bibitem[{\citenamefont{Horodecki et~al.}(2009)\citenamefont{Horodecki,
  Horodecki, Horodecki, and Horodecki}}]{horodecki2009quantum}
\bibinfo{author}{\bibfnamefont{R.}~\bibnamefont{Horodecki}},
  \bibinfo{author}{\bibfnamefont{P.}~\bibnamefont{Horodecki}},
  \bibinfo{author}{\bibfnamefont{M.}~\bibnamefont{Horodecki}},
  \bibnamefont{and}
  \bibinfo{author}{\bibfnamefont{K.}~\bibnamefont{Horodecki}},
  \bibinfo{journal}{Rev. Mod. Phys.} \textbf{\bibinfo{volume}{81}},
  \bibinfo{pages}{865} (\bibinfo{year}{2009}).

\bibitem[{\citenamefont{Bose et~al.}(1998)\citenamefont{Bose, Vedral, and
  Knight}}]{bose1998multiparticle}
\bibinfo{author}{\bibfnamefont{S.}~\bibnamefont{Bose}},
  \bibinfo{author}{\bibfnamefont{V.}~\bibnamefont{Vedral}}, \bibnamefont{and}
  \bibinfo{author}{\bibfnamefont{P.~L.} \bibnamefont{Knight}},
  \bibinfo{journal}{Phys. Rev. A} \textbf{\bibinfo{volume}{57}},
  \bibinfo{pages}{822} (\bibinfo{year}{1998}).

\bibitem[{\citenamefont{Goebel et~al.}(2008)\citenamefont{Goebel, Wagenknecht,
  Zhang, Chen, Chen, Schmiedmayer, and Pan}}]{2008multistage}
\bibinfo{author}{\bibfnamefont{A.~M.} \bibnamefont{Goebel}},
  \bibinfo{author}{\bibfnamefont{C.}~\bibnamefont{Wagenknecht}},
  \bibinfo{author}{\bibfnamefont{Q.}~\bibnamefont{Zhang}},
  \bibinfo{author}{\bibfnamefont{Y.-A.} \bibnamefont{Chen}},
  \bibinfo{author}{\bibfnamefont{K.}~\bibnamefont{Chen}},
  \bibinfo{author}{\bibfnamefont{J.}~\bibnamefont{Schmiedmayer}},
  \bibnamefont{and} \bibinfo{author}{\bibfnamefont{J.-W.} \bibnamefont{Pan}},
  \bibinfo{journal}{Phys. Rev. Lett.} \textbf{\bibinfo{volume}{101}},
  \bibinfo{pages}{080403} (\bibinfo{year}{2008}).

\bibitem[{\citenamefont{Khalique et~al.}(2013)\citenamefont{Khalique, Tittel,
  and Sanders}}]{2013multistage}
\bibinfo{author}{\bibfnamefont{A.}~\bibnamefont{Khalique}},
  \bibinfo{author}{\bibfnamefont{W.}~\bibnamefont{Tittel}}, \bibnamefont{and}
  \bibinfo{author}{\bibfnamefont{B.~C.} \bibnamefont{Sanders}},
  \bibinfo{journal}{Phys. Rev. A} \textbf{\bibinfo{volume}{88}},
  \bibinfo{pages}{022336} (\bibinfo{year}{2013}).

\bibitem[{\citenamefont{Lu et~al.}(2009)\citenamefont{Lu, Yang, and
  Pan}}]{MultipartEnt2009}
\bibinfo{author}{\bibfnamefont{C.-Y.} \bibnamefont{Lu}},
  \bibinfo{author}{\bibfnamefont{T.}~\bibnamefont{Yang}}, \bibnamefont{and}
  \bibinfo{author}{\bibfnamefont{J.-W.} \bibnamefont{Pan}},
  \bibinfo{journal}{Phys. Rev. Lett.} \textbf{\bibinfo{volume}{103}},
  \bibinfo{pages}{020501} (\bibinfo{year}{2009}).

\bibitem[{\citenamefont{Vamivakas et~al.}(2004)\citenamefont{Vamivakas, Saleh,
  Sergienko, and Teich}}]{SPDC}
\bibinfo{author}{\bibfnamefont{A.~N.} \bibnamefont{Vamivakas}},
  \bibinfo{author}{\bibfnamefont{B.~E.} \bibnamefont{Saleh}},
  \bibinfo{author}{\bibfnamefont{A.~V.} \bibnamefont{Sergienko}},
  \bibnamefont{and} \bibinfo{author}{\bibfnamefont{M.~C.} \bibnamefont{Teich}},
  \bibinfo{journal}{Phys. Rev. A} \textbf{\bibinfo{volume}{70}},
  \bibinfo{pages}{043810} (\bibinfo{year}{2004}).

\bibitem[{\citenamefont{Dousse et~al.}(2010)\citenamefont{Dousse,
  Suffczy{\'n}ski, Beveratos, Krebs, Lema{\^\i}tre, Sagnes, Bloch, Voisin, and
  Senellart}}]{SPDC2}
\bibinfo{author}{\bibfnamefont{A.}~\bibnamefont{Dousse}},
  \bibinfo{author}{\bibfnamefont{J.}~\bibnamefont{Suffczy{\'n}ski}},
  \bibinfo{author}{\bibfnamefont{A.}~\bibnamefont{Beveratos}},
  \bibinfo{author}{\bibfnamefont{O.}~\bibnamefont{Krebs}},
  \bibinfo{author}{\bibfnamefont{A.}~\bibnamefont{Lema{\^\i}tre}},
  \bibinfo{author}{\bibfnamefont{I.}~\bibnamefont{Sagnes}},
  \bibinfo{author}{\bibfnamefont{J.}~\bibnamefont{Bloch}},
  \bibinfo{author}{\bibfnamefont{P.}~\bibnamefont{Voisin}}, \bibnamefont{and}
  \bibinfo{author}{\bibfnamefont{P.}~\bibnamefont{Senellart}},
  \bibinfo{journal}{Nature} \textbf{\bibinfo{volume}{466}},
  \bibinfo{pages}{217} (\bibinfo{year}{2010}).

\bibitem[{\citenamefont{Scherer et~al.}(2009)\citenamefont{Scherer, Howard,
  Sanders, and Tittel}}]{problemsSPDC}
\bibinfo{author}{\bibfnamefont{A.}~\bibnamefont{Scherer}},
  \bibinfo{author}{\bibfnamefont{R.~B.} \bibnamefont{Howard}},
  \bibinfo{author}{\bibfnamefont{B.~C.} \bibnamefont{Sanders}},
  \bibnamefont{and} \bibinfo{author}{\bibfnamefont{W.}~\bibnamefont{Tittel}},
  \bibinfo{journal}{Phys. Rev. A} \textbf{\bibinfo{volume}{80}},
  \bibinfo{pages}{062310} (\bibinfo{year}{2009}).

\bibitem[{\citenamefont{Lee and Jeong}(2013)}]{lee2013bell}
\bibinfo{author}{\bibfnamefont{S.-W.} \bibnamefont{Lee}} \bibnamefont{and}
  \bibinfo{author}{\bibfnamefont{H.}~\bibnamefont{Jeong}},
  \bibinfo{journal}{Proceedings of the First International Conference on Entangled Coherent State and Its Application to Quantum Information Science, Tamagawa University, Tokyo},
  \bibinfo{pages}{41-46} (\bibinfo{year}{2012}).

  
 


\bibitem[{\citenamefont{Yang et~al.}(2005)\citenamefont{Yang, Song, and
  Cao}}]{cQED}
\bibinfo{author}{\bibfnamefont{M.}~\bibnamefont{Yang}},
  \bibinfo{author}{\bibfnamefont{W.}~\bibnamefont{Song}}, \bibnamefont{and}
  \bibinfo{author}{\bibfnamefont{Z.-L.} \bibnamefont{Cao}},
  \bibinfo{journal}{Phys. Rev. A} \textbf{\bibinfo{volume}{71}},
  \bibinfo{pages}{034312} (\bibinfo{year}{2005}).

\bibitem[{\citenamefont{Pakniat
  et~al.}(2017{\natexlab{a}})\citenamefont{Pakniat, Zandi, and
  Tavassoly}}]{cQED3}
\bibinfo{author}{\bibfnamefont{R.}~\bibnamefont{Pakniat}},
  \bibinfo{author}{\bibfnamefont{M.~H.} \bibnamefont{Zandi}}, \bibnamefont{and}
  \bibinfo{author}{\bibfnamefont{M.~K.} \bibnamefont{Tavassoly}},
  \bibinfo{journal}{Eur. Phys. J. Plus} \textbf{\bibinfo{volume}{132}},
  \bibinfo{pages}{3} (\bibinfo{year}{2017}{\natexlab{a}}).

\bibitem[{\citenamefont{Grice}(2011)}]{PhysRevA.84.042331}
\bibinfo{author}{\bibfnamefont{W.~P.} \bibnamefont{Grice}},
  \bibinfo{journal}{Phys. Rev. A} \textbf{\bibinfo{volume}{84}},
  \bibinfo{pages}{042331} (\bibinfo{year}{2011}).

\bibitem[{\citenamefont{Zhou and Sheng}(2015)}]{PhysRevA.92.042314}
\bibinfo{author}{\bibfnamefont{L.}~\bibnamefont{Zhou}} \bibnamefont{and}
  \bibinfo{author}{\bibfnamefont{Y.-B.} \bibnamefont{Sheng}},
  \bibinfo{journal}{Phys. Rev. A} \textbf{\bibinfo{volume}{92}},
  \bibinfo{pages}{042314} (\bibinfo{year}{2015}).

\bibitem[{\citenamefont{Sheng et~al.}(2010)\citenamefont{Sheng, Deng, and
  Long}}]{PhysRevA.82.032318}
\bibinfo{author}{\bibfnamefont{Y.-B.} \bibnamefont{Sheng}},
  \bibinfo{author}{\bibfnamefont{F.-G.} \bibnamefont{Deng}}, \bibnamefont{and}
  \bibinfo{author}{\bibfnamefont{G.~L.} \bibnamefont{Long}},
  \bibinfo{journal}{Phys. Rev. A} \textbf{\bibinfo{volume}{82}},
  \bibinfo{pages}{032318} (\bibinfo{year}{2010}).

\bibitem[{\citenamefont{Pirandola et~al.}(2006)\citenamefont{Pirandola, Vitali,
  Tombesi, and Lloyd}}]{macroscES2006}
\bibinfo{author}{\bibfnamefont{S.}~\bibnamefont{Pirandola}},
  \bibinfo{author}{\bibfnamefont{D.}~\bibnamefont{Vitali}},
  \bibinfo{author}{\bibfnamefont{P.}~\bibnamefont{Tombesi}}, \bibnamefont{and}
  \bibinfo{author}{\bibfnamefont{S.}~\bibnamefont{Lloyd}},
  \bibinfo{journal}{Phys. Rev. Lett.} \textbf{\bibinfo{volume}{97}},
  \bibinfo{pages}{150403} (\bibinfo{year}{2006}).

\bibitem[{\citenamefont{Press et~al.}(2008)\citenamefont{Press, Ladd, Zhang,
  and Yamamoto}}]{Yamamoto2008}
\bibinfo{author}{\bibfnamefont{D.}~\bibnamefont{Press}},
  \bibinfo{author}{\bibfnamefont{T.~D.} \bibnamefont{Ladd}},
  \bibinfo{author}{\bibfnamefont{B.}~\bibnamefont{Zhang}}, \bibnamefont{and}
  \bibinfo{author}{\bibfnamefont{Y.}~\bibnamefont{Yamamoto}},
  \bibinfo{journal}{Nature} \textbf{\bibinfo{volume}{456}},
  \bibinfo{pages}{218} (\bibinfo{year}{2008}).

\bibitem[{\citenamefont{F{\`e}ve et~al.}(2007)\citenamefont{F{\`e}ve, Mah{\'e},
  Berroir, Kontos, Placais, Glattli, Cavanna, Etienne, and
  Jin}}]{feve2007demand}
\bibinfo{author}{\bibfnamefont{G.}~\bibnamefont{F{\`e}ve}},
  \bibinfo{author}{\bibfnamefont{A.}~\bibnamefont{Mah{\'e}}},
  \bibinfo{author}{\bibfnamefont{J.-M.} \bibnamefont{Berroir}},
  \bibinfo{author}{\bibfnamefont{T.}~\bibnamefont{Kontos}},
  \bibinfo{author}{\bibfnamefont{B.}~\bibnamefont{Placais}},
  \bibinfo{author}{\bibfnamefont{D.}~\bibnamefont{Glattli}},
  \bibinfo{author}{\bibfnamefont{A.}~\bibnamefont{Cavanna}},
  \bibinfo{author}{\bibfnamefont{B.}~\bibnamefont{Etienne}}, \bibnamefont{and}
  \bibinfo{author}{\bibfnamefont{Y.}~\bibnamefont{Jin}},
  \bibinfo{journal}{Science} \textbf{\bibinfo{volume}{316}},
  \bibinfo{pages}{1169} (\bibinfo{year}{2007}).

\bibitem[{\citenamefont{Bocquillon et~al.}(2013)\citenamefont{Bocquillon,
  Freulon, Berroir, Degiovanni, Pla{\c{c}}ais, Cavanna, Jin, and
  F{\`e}ve}}]{bocquillon2013coherence}
\bibinfo{author}{\bibfnamefont{E.}~\bibnamefont{Bocquillon}},
  \bibinfo{author}{\bibfnamefont{V.}~\bibnamefont{Freulon}},
  \bibinfo{author}{\bibfnamefont{J.-M.} \bibnamefont{Berroir}},
  \bibinfo{author}{\bibfnamefont{P.}~\bibnamefont{Degiovanni}},
  \bibinfo{author}{\bibfnamefont{B.}~\bibnamefont{Pla{\c{c}}ais}},
  \bibinfo{author}{\bibfnamefont{A.}~\bibnamefont{Cavanna}},
  \bibinfo{author}{\bibfnamefont{Y.}~\bibnamefont{Jin}}, \bibnamefont{and}
  \bibinfo{author}{\bibfnamefont{G.}~\bibnamefont{F{\`e}ve}},
  \bibinfo{journal}{Science} \textbf{\bibinfo{volume}{339}},
  \bibinfo{pages}{1054} (\bibinfo{year}{2013}).

\bibitem[{\citenamefont{Rashidi et~al.}(2018)\citenamefont{Rashidi, Vine,
  Dienel, Livadaru, Retallick, Huff, Walus, and
  Wolkow}}]{PhysRevLett.121.166801}
\bibinfo{author}{\bibfnamefont{M.}~\bibnamefont{Rashidi}},
  \bibinfo{author}{\bibfnamefont{W.}~\bibnamefont{Vine}},
  \bibinfo{author}{\bibfnamefont{T.}~\bibnamefont{Dienel}},
  \bibinfo{author}{\bibfnamefont{L.}~\bibnamefont{Livadaru}},
  \bibinfo{author}{\bibfnamefont{J.}~\bibnamefont{Retallick}},
  \bibinfo{author}{\bibfnamefont{T.}~\bibnamefont{Huff}},
  \bibinfo{author}{\bibfnamefont{K.}~\bibnamefont{Walus}}, \bibnamefont{and}
  \bibinfo{author}{\bibfnamefont{R.~A.} \bibnamefont{Wolkow}},
  \bibinfo{journal}{Phys. Rev. Lett.} \textbf{\bibinfo{volume}{121}},
  \bibinfo{pages}{166801} (\bibinfo{year}{2018}).

\bibitem[{\citenamefont{Matthews et~al.}(2013)\citenamefont{Matthews, Poulios,
  Meinecke, Politi, Peruzzo, Ismail, W{\"o}rhoff, Thompson, and
  O'Brien}}]{FermionicStatistics}
\bibinfo{author}{\bibfnamefont{J.~C.} \bibnamefont{Matthews}},
  \bibinfo{author}{\bibfnamefont{K.}~\bibnamefont{Poulios}},
  \bibinfo{author}{\bibfnamefont{J.~D.} \bibnamefont{Meinecke}},
  \bibinfo{author}{\bibfnamefont{A.}~\bibnamefont{Politi}},
  \bibinfo{author}{\bibfnamefont{A.}~\bibnamefont{Peruzzo}},
  \bibinfo{author}{\bibfnamefont{N.}~\bibnamefont{Ismail}},
  \bibinfo{author}{\bibfnamefont{K.}~\bibnamefont{W{\"o}rhoff}},
  \bibinfo{author}{\bibfnamefont{M.~G.} \bibnamefont{Thompson}},
  \bibnamefont{and} \bibinfo{author}{\bibfnamefont{J.~L.}
  \bibnamefont{O'Brien}}, \bibinfo{journal}{Sci. Rep.}
  \textbf{\bibinfo{volume}{3}}, \bibinfo{pages}{1539} (\bibinfo{year}{2013}).

\bibitem[{\citenamefont{Sansoni et~al.}(2012)\citenamefont{Sansoni, Sciarrino,
  Vallone, Mataloni, Crespi, Ramponi, and Osellame}}]{sansoni2012two}
\bibinfo{author}{\bibfnamefont{L.}~\bibnamefont{Sansoni}},
  \bibinfo{author}{\bibfnamefont{F.}~\bibnamefont{Sciarrino}},
  \bibinfo{author}{\bibfnamefont{G.}~\bibnamefont{Vallone}},
  \bibinfo{author}{\bibfnamefont{P.}~\bibnamefont{Mataloni}},
  \bibinfo{author}{\bibfnamefont{A.}~\bibnamefont{Crespi}},
  \bibinfo{author}{\bibfnamefont{R.}~\bibnamefont{Ramponi}}, \bibnamefont{and}
  \bibinfo{author}{\bibfnamefont{R.}~\bibnamefont{Osellame}},
  \bibinfo{journal}{Phys. Rev. Lett.} \textbf{\bibinfo{volume}{108}},
  \bibinfo{pages}{010502} (\bibinfo{year}{2012}).

\bibitem[{\citenamefont{Orzel et~al.}(1999)\citenamefont{Orzel, Walhout, Sterr,
  Julienne, and Rolston}}]{orzelPRA}
\bibinfo{author}{\bibfnamefont{C.}~\bibnamefont{Orzel}},
  \bibinfo{author}{\bibfnamefont{M.}~\bibnamefont{Walhout}},
  \bibinfo{author}{\bibfnamefont{U.}~\bibnamefont{Sterr}},
  \bibinfo{author}{\bibfnamefont{P.~S.} \bibnamefont{Julienne}},
  \bibnamefont{and} \bibinfo{author}{\bibfnamefont{S.~L.}
  \bibnamefont{Rolston}}, \bibinfo{journal}{Phys. Rev. A}
  \textbf{\bibinfo{volume}{59}}, \bibinfo{pages}{1926} (\bibinfo{year}{1999}).

\bibitem[{\citenamefont{Plenio and Virmani}(2007)}]{plenio2007mb}
\bibinfo{author}{\bibfnamefont{M.}~\bibnamefont{Plenio}} \bibnamefont{and}
  \bibinfo{author}{\bibfnamefont{S.}~\bibnamefont{Virmani}},
  \bibinfo{journal}{Quantum Inf. Comput.} \textbf{\bibinfo{volume}{7}},
  \bibinfo{pages}{1} (\bibinfo{year}{2007}).

\bibitem[{\citenamefont{Buks et~al.}(1998)\citenamefont{Buks, Schuster,
  Heiblum, Mahalu, and Umansky}}]{detectorselectrons}
\bibinfo{author}{\bibfnamefont{E.}~\bibnamefont{Buks}},
  \bibinfo{author}{\bibfnamefont{R.}~\bibnamefont{Schuster}},
  \bibinfo{author}{\bibfnamefont{M.}~\bibnamefont{Heiblum}},
  \bibinfo{author}{\bibfnamefont{D.}~\bibnamefont{Mahalu}}, \bibnamefont{and}
  \bibinfo{author}{\bibfnamefont{V.}~\bibnamefont{Umansky}},
  \bibinfo{journal}{Nature} \textbf{\bibinfo{volume}{391}},
  \bibinfo{pages}{871} (\bibinfo{year}{1998}).

\bibitem[{\citenamefont{Armour and Blencowe}(2001)}]{detectors}
\bibinfo{author}{\bibfnamefont{A.~D.} \bibnamefont{Armour}} \bibnamefont{and}
  \bibinfo{author}{\bibfnamefont{M.~P.} \bibnamefont{Blencowe}},
  \bibinfo{journal}{Phys. Rev. B} \textbf{\bibinfo{volume}{64}},
  \bibinfo{pages}{035311} (\bibinfo{year}{2001}).

\bibitem[{\citenamefont{Parkins et~al.}(1995)\citenamefont{Parkins, Marte,
  Zoller, Carnal, and Kimble}}]{detectors2}
\bibinfo{author}{\bibfnamefont{A.~S.} \bibnamefont{Parkins}},
  \bibinfo{author}{\bibfnamefont{P.}~\bibnamefont{Marte}},
  \bibinfo{author}{\bibfnamefont{P.}~\bibnamefont{Zoller}},
  \bibinfo{author}{\bibfnamefont{O.}~\bibnamefont{Carnal}}, \bibnamefont{and}
  \bibinfo{author}{\bibfnamefont{H.~J.} \bibnamefont{Kimble}},
  \bibinfo{journal}{Phys. Rev. A} \textbf{\bibinfo{volume}{51}},
  \bibinfo{pages}{1578} (\bibinfo{year}{1995}).

\bibitem[{\citenamefont{Xiu et~al.}(2007)\citenamefont{Xiu, Hong-Cai, Rong-Can,
  and Zhi-Ping}}]{cQED2}
\bibinfo{author}{\bibfnamefont{L.}~\bibnamefont{Xiu}},
  \bibinfo{author}{\bibfnamefont{L.}~\bibnamefont{Hong-Cai}},
  \bibinfo{author}{\bibfnamefont{Y.}~\bibnamefont{Rong-Can}}, \bibnamefont{and}
  \bibinfo{author}{\bibfnamefont{H.}~\bibnamefont{Zhi-Ping}},
  \bibinfo{journal}{Chin. Phys.} \textbf{\bibinfo{volume}{16}},
  \bibinfo{pages}{919} (\bibinfo{year}{2007}).

\bibitem[{\citenamefont{Pakniat
  et~al.}(2017{\natexlab{b}})\citenamefont{Pakniat, Tavassoly, and
  Zandi}}]{CQED2017}
\bibinfo{author}{\bibfnamefont{R.}~\bibnamefont{Pakniat}},
  \bibinfo{author}{\bibfnamefont{M.}~\bibnamefont{Tavassoly}},
  \bibnamefont{and} \bibinfo{author}{\bibfnamefont{M.}~\bibnamefont{Zandi}},
  \bibinfo{journal}{Opt. Comm.} \textbf{\bibinfo{volume}{382}},
  \bibinfo{pages}{381} (\bibinfo{year}{2017}{\natexlab{b}}).

\bibitem[{\citenamefont{B{\"a}uerle et~al.}(2018)\citenamefont{B{\"a}uerle,
  Glattli, Meunier, Portier, Roche, Roulleau, Takada, and
  Waintal}}]{electronsReview}
\bibinfo{author}{\bibfnamefont{C.}~\bibnamefont{B{\"a}uerle}},
  \bibinfo{author}{\bibfnamefont{D.~C.} \bibnamefont{Glattli}},
  \bibinfo{author}{\bibfnamefont{T.}~\bibnamefont{Meunier}},
  \bibinfo{author}{\bibfnamefont{F.}~\bibnamefont{Portier}},
  \bibinfo{author}{\bibfnamefont{P.}~\bibnamefont{Roche}},
  \bibinfo{author}{\bibfnamefont{P.}~\bibnamefont{Roulleau}},
  \bibinfo{author}{\bibfnamefont{S.}~\bibnamefont{Takada}}, \bibnamefont{and}
  \bibinfo{author}{\bibfnamefont{X.}~\bibnamefont{Waintal}},
  \bibinfo{journal}{Rep. Prog. Phys.} \textbf{\bibinfo{volume}{81}},
  \bibinfo{pages}{056503} (\bibinfo{year}{2018}).

\end{thebibliography}

\end{document}